# XMDCA-TL: An Explainable Multi-Domain Channel Attention Transfer Learning Framework for Fault Diagnosis in Industrial Gas Turbines

Amir Jahangard Takaloo - *Electrical Engineering Department* - K.N. Toosi University of Technology, Tehran, IRAN

amir.jahangardtakaloo@email.kntu.ac.ir

Mahdi Aliyari Shoorehdeli* - *Electrical Engineering Department* - K.N. Toosi University of Technology, Tehran, IRAN

aliyari@kntu.ac.ir

Ehsan Mohammadi - *Digital Technology Development Dept.*, - MAPNA Digital Co., Tehran, IRAN

Mohammadi_eh@mapnagroup.com

* Corresponding author. Email address: aliyari@kntu.ac.ir

**Abstract**

In this study, a multi-task, interpretable transfer learning framework, XMDCA-TL, is proposed for fault diagnosis in industrial gas turbines. In the proposed method, the vibration time waveform is first converted into a multi-domain RGB representation comprising time, frequency, and time-frequency domains. A ConvNeXtV2-based encoder then processes these images, and the Multi-Domain Channel Attention (MDCA) mechanism is applied to its deep layers to model interactions among different domains and complementary dependencies in the signals. To improve the quality of the learned representations and enhance the model's robustness to noise, a self-supervised strategy based on hybrid masking, along with a UNet-based decoder to reconstruct the masked regions, has been designed. To overcome the limitation of labeled data in industrial environments, transfer learning was employed to transfer knowledge from laboratory data to real-world data from a 42.2 MW MGT-40 gas turbine at the Zahedan power plant. Additionally, a comprehensive Explainable Artificial Intelligence (XAI) framework was developed to analyze decision-making regions, evaluate domain importance, examine the flow of attention between domains, and assess reconstruction uncertainty. The results showed that XMDCA-TL, while achieving satisfactory fault diagnosis performance, possesses domain adaptability and robustness to noise and provides a physical interpretation of the model's decision-making process.

## 1. Introduction

Rotating machinery is among the most important mechanical equipment used across various industries, including power plants, petrochemicals, steel, and production lines [1]. Among this equipment, industrial gas turbines are considered among the most important applications. Its proper operation plays an important role in the stability of industrial processes and in reducing maintenance costs [2]. However, over time, these assets are affected by factors such as misalignment, imbalance, and bearing faults. This can lead to reduced efficiency, sudden system shutdowns, and significant economic losses [3].

In this context, vibration signals are widely recognized as one of the most effective sources of information for assessing the health status of rotating machinery [4]. Despite significant progress in this field, traditional methods based on manual feature extraction rely on expert knowledge and specific operating conditions [5]. Thus, they suffer from performance degradation under changing conditions and with limited labeled data. However, deep learning methods provide powerful automatic feature extraction but often require large amounts of labeled data [2]. These limitations have paved the way for the emergence of novel, effective approaches to improve model generalization and optimize the use of available data, such as Transfer Learning (TL) and Self-Supervised Learning (SSL).

Despite recent advances, feature extraction techniques for vibration signals are highly varied. In many industrial applications, features are extracted directly from the time and frequency domains and used as one-dimensional signals. More recent studies have enabled the use of machine vision networks by transforming vibration signals into two-dimensional representations. These imaging approaches can more effectively reveal spatial patterns and hidden relationships among different signal representations [6, 7, 8]. However, a significant gap still exists between signal analysis and the capabilities of modern deep learning models. For this reason, the goal of this study is to integrate vibration analysis expertise with modern deep learning tools for signal processing, using transfer learning and explainable deep learning techniques.

In recent years, graphical representations of vibration signals have been widely adopted [9]. However, existing methods often focus on a single analytical domain, which may prevent the full exploitation of the information contained in the signal [10, 11]. Moreover, converting vibration signals into multidimensional visual representations has become an effective approach for applying deep learning models to rotating machinery fault diagnosis [12]. The main reason is that each analytical domain reveals only a portion of the information embedded in the signal. Unfortunately, a single representation usually cannot fully describe the system's dynamic behavior. In this study, a three-domain RGB representation is employed. Hence, the model will be able to extract more meaningful feature representations of the equipment's health status, thereby improving its fault-diagnosis performance.

In rotating machinery systems, measured vibration signals are affected by ambient noise, sensor errors, load variations, and speed fluctuations. These factors reduce the Signal-to-Noise Ratio (SNR), making it more difficult to extract features related to the fault [13]. Thus, developing models that can reconstruct incomplete or corrupted information has become particularly important [14]. Deep learning-based approaches, unlike classical filters, can learn the complex nonlinear structures of vibration data and produce more robust representations [1]. In many industrial applications, acquiring labeled data requires costly experiments and expert involvement. However, vast amounts of unlabeled data are generated daily by monitoring systems [15]. This reality has positioned Self-Supervised Learning (SSL) as a novel approach to representation learning. In which a neural network solves a pretext task and learns the intrinsic structure of the data. Here, a masking strategy is employed as the pretext task.

The proposed architecture in this study, referred to as XMDCA-TL (Explainable Multi-Domain Channel Attention Transfer Learning), is based on an explainable multitask transfer learning framework. As shown in Figure 1, it comprises two main parts: an encoder based on ConvNeXtV2, equipped with a multi-domain channel attention (MDCA) mechanism, and a UNet-type decoder.

In the proposed framework, the encoder's backbone is designed based on the ConvNeXtV2 architecture [16]. This architecture is an optimized variant of modern convolutional networks, inspired by the Vision Transformer's structure. The RGB images generated from vibration signals differ in nature from those of ordinary images. Although popular CNN models can extract local features, they lack a dedicated mechanism to explicitly model cross-domain dependencies. As a result, the backbone implicitly attempts to learn these interactions, potentially leading to the loss of important physical information. To address this limitation, a Multi-Domain Channel Attention (MDCA) mechanism is designed that directly models the mutual dependencies among the three physical domains. Moreover, this framework enables interpretable analysis of each signal domain's contribution and examination of the cross-domain interactions learned during feature extraction.

Alongside the encoder, a denoising decoder based on the UNet architecture is employed to reconstruct corrupted and incomplete data. Since vibration signals in industrial environments are typically affected by noise and disturbances, relying solely on extracted features for classification may be inadequate. For this reason, the decoder uses skip connections to leverage the encoder's multi-scale features. In addition, the primary task of fault diagnosis is entrusted to a lightweight, purpose-built classification head that follows immediately after the MDCA modules. The goal of this block is to convert the high-level representations extracted by the encoder into the final fault class label.

One of the key aspects of this research is the combination of transfer learning with the MDCA mechanism. Since the pretrained weights have been primarily trained on natural RGB images, applying them directly to vibrational RGB data may not be optimal. In this regard, MDCA acts as a bridge between the general pretrained knowledge and the specialized structure of the target domain. Using transfer learning in this architecture brings significant performance advantages. In fact, the need for labeled data is drastically reduced, training speed increases, and the likelihood of overfitting decreases. Moreover, the model produces representations that are more robust under variable conditions. This characteristic is especially important in industrial applications, where real-world data often includes noise, changing operating conditions, sensor variations, and domain shift [17].

Despite the remarkable performance of deep learning models in rotating machinery fault diagnosis, a fundamental challenge for these methods is their black-box nature. In many industrial applications, it is necessary to understand which parts of the data the model relied on to make its decision. As a result, the use of Explainable Artificial Intelligence (XAI) methods has gained widespread attention for enhancing the transparency, trustworthiness, and interpretability of models. These methods enable the identification of features, regions, and patterns that most strongly influence the network's decision-making. In other words, they reveal the connection between the model's output and the system's behavior [18, 19]. In this study, interpretability analysis is not limited to identifying important regions of the signal. A multi-level XAI framework has indeed been designed to examine the model's behavior at different levels. This multi-layered approach enables a physical interpretation of the fault diagnosis process and of how the network utilizes vibrational information.

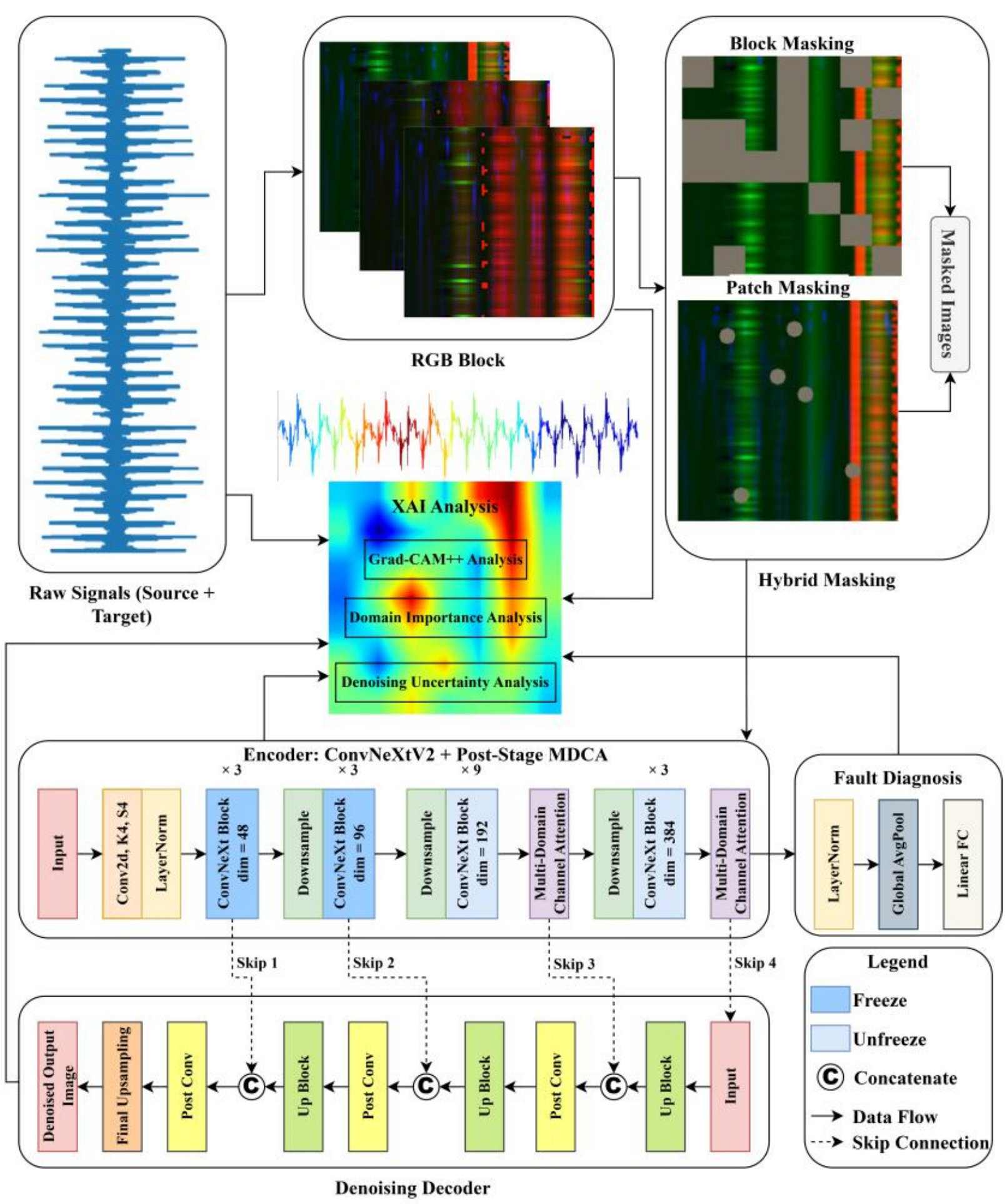

*Figure 1: Architecture of the Proposed XMDCA-TL Framework.*

In recent years, the use of transfer learning based on deep vision networks for the fault diagnosis of rotating machinery has attracted attention. For example, a ConvNeXt-based transfer learning method called TL-CoCNN was proposed, in which vibration signals were converted into RGB images and used to classify faults. Their results showed that combining a multi-domain signal representation with transfer learning can improve diagnosis accuracy. However, this method does not provide a specific mechanism for interpreting the network's decision-making process [12].

A deep transfer learning framework called Supervised Dynamic Sensor Scoring (SDSS) was introduced for interpretable fault diagnosis in industrial gas turbines. The results showed that the SDSS framework, in addition to improving fault diagnosis performance under labeled data scarcity, can be an effective step toward developing interpretable deep learning-based fault diagnosis systems for industrial applications [20]. As another example, a method called SCMGIDTL for unsupervised fault diagnosis of rotating machinery was proposed. Experimental results showed that this approach can extract transferable knowledge from multiple source domains and achieve very high performance in unsupervised fault diagnosis of rotating machinery under different operating conditions [21].

In the last few years, the combination of transfer learning and attention-based models has become a key approach to fault diagnosis. For example, a transformer-based framework called IDAT was introduced. The results of this study indicate that using attention mechanisms can not only improve transfer learning performance but also enable the analysis and interpretation of the model's decision-making process [17]. In an effort to combine transfer learning and explainable AI for rotating machine fault diagnosis, a framework called FaultD-XAI was developed, in which synthetic data generated for model training and knowledge transfer were applied to real data. The results of this study showed that the combination of synthetic data, transfer learning, and XAI methods can reduce reliance on real labeled data while increasing the reliability of fault diagnosis [18].

The most significant innovations and contributions of this research can be summarized as follows:

- Presenting a multi-task and interpretable transfer learning framework for rotating machinery fault diagnosis that simultaneously performs fault diagnosis, self-supervised data reconstruction, and model interpretability analysis.
- Introducing a multi-domain RGB representation based on information from the time, frequency, and time-frequency domains, along with a multi-domain channel attention mechanism to leverage complementary information and model interactions among different domains.
- Designing a self-supervised noise removal strategy based on hybrid masking and a UNet-based decoder to improve representation learning, enhance noise robustness, and facilitate cross-domain consistency.
- Development of a comprehensive XAI framework for analyzing model behavior at the signal and domain levels, including identifying regions influential in decision-making, assessing domain importance, analyzing attention flow between domains, and examining patterns of uncertainty and data reconstruction.

The remainder of this paper starts with the methodology and explains the proposed method in Section 2. The dataset is introduced in Section 3, and the verification and results of the proposed algorithm are presented in Section 4. In the end, Section 5 provides a conclusion of this paper.

## 2. Methodology

### 2.1. Data Visualization

In this study, to extract complementary information from vibration signals, a multi-domain structure based on RGB images is used. The main idea of this method is that each analytical domain describes only part of the system's physical behavior. In other words, no single domain can fully capture the fault's dynamic characteristics. For this reason, three different representations – time domain, frequency domain, and time-frequency domain – are extracted and then combined as three independent color channels in an RGB image. In this structure, the red (R) channel represents the time domain, the green (G) channel represents the frequency domain, and the blue (B) channel displays the time-frequency domain. This combination allows the deep learning model to simultaneously observe and learn temporal interactions, spectral structures, and non-stationary changes.

First, the raw vibration signal is divided into smaller segments to extract local features. If we denote the original signal by x[n], each segment of fixed length L is defined as follows:

$$x_k[n] = x[n + k.s], 0 \leq n < L \tag{1}$$

where s is the stride between two consecutive segments. Using segmentation increases the number of training samples, better extracts local patterns, and reduces the model's sensitivity to random variations. In this study, each segment is set to 4096 samples, and overlap is achieved with a stride of 512. This overlap ensures that transient information and short-term signal variations are not lost. The flowchart for converting vibration time-series data into an RGB image is shown in Figure 2. A more detailed examination of each channel is provided below.

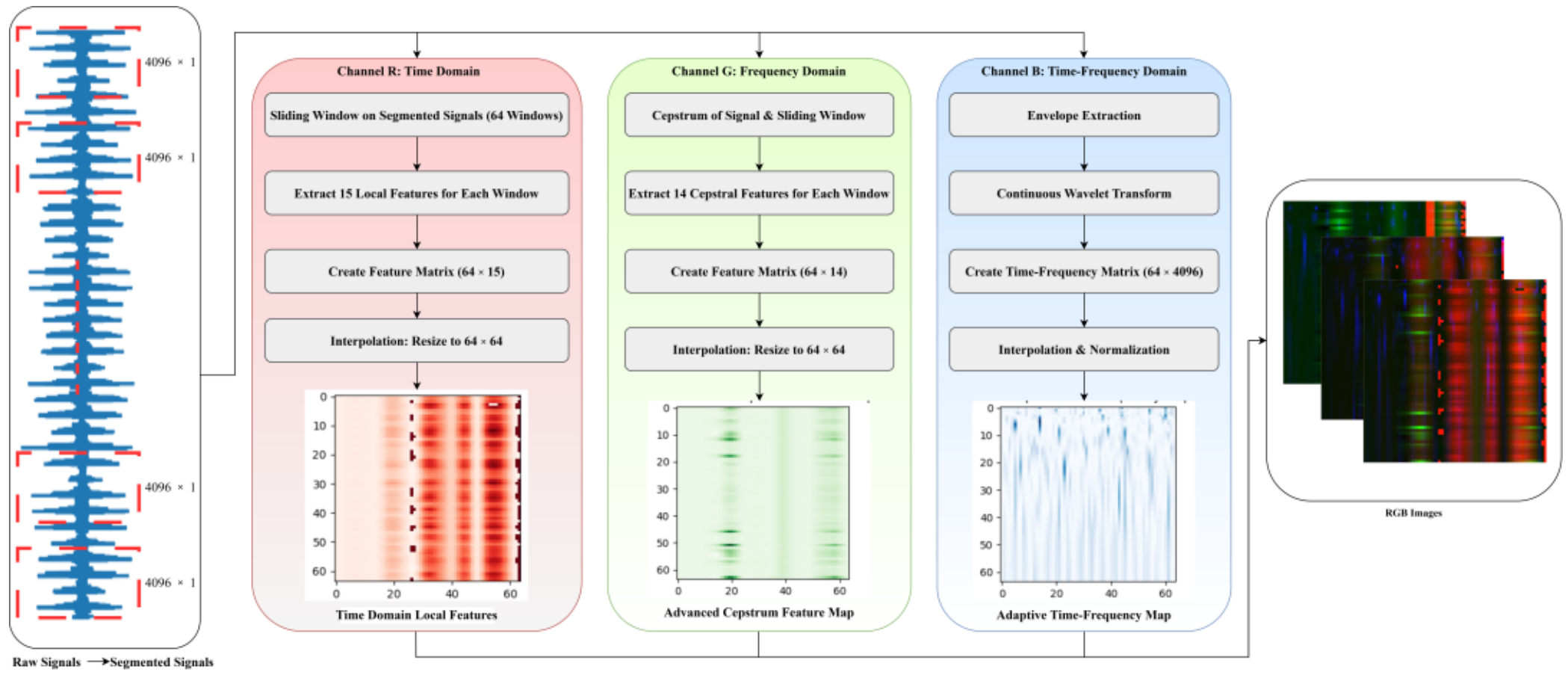

*Figure 2: Workflow of Vibration Signal Conversion into Multi-Domain RGB Images.*

### 2.1.1. Channel R

In this study, the red channel of the RGB image is used to extract and display time-domain features. The primary goal of this channel's design is to preserve the direct, raw behavior of the vibrational signal and to extract local patterns. Time-domain analysis directly examines amplitude variations, impacts, sudden fluctuations, transient behaviors, and statistical changes in the signal. Many mechanical faults – such as imbalance, bearing faults, misalignment, and looseness – initially manifest as localized variations in the time domain [22]. In the proposed method, segmented signals are divided into smaller windows with specified overlap to extract local behavior. A window length of 128 samples and a stride of 64 were chosen. Hence, 64 windows have been created.

Then, for each window, 15 statistical and dynamic features are extracted. These features include Mean, Standard Deviation, Mean Absolute Deviation, RMS Value, Peak-to-Peak Amplitude, Maximum Absolute Value, Skewness, Kurtosis, Crest Factor, Shape Factor, Impulse Factor, Modified Impulse Factor, Clearance Factor, Form Factor, and Signal Energy. Each of these features describes a different aspect of the physical behavior of vibration. For example, RMS indicates the overall vibration energy, while kurtosis is highly sensitive to impacts and impulsive behavior and is commonly used in bearing fault diagnosis.

After extracting features from all windows, a two-dimensional feature matrix is formed, with rows corresponding to time windows and columns to extracted features. Thus, the feature matrix is 64×15. Next, the feature values are normalized using Min-Max normalization so that all features lie in the range [0,1]. Then, the data are mapped to the image range ([0,255]) to allow them to be stored as a grayscale image. Since the initial feature matrix does not match the network's required dimensions, 2D interpolation is used to resize it to 64×64. For this purpose, the cubic interpolation method in OpenCV is used [23, 24]. Ultimately, the output of this stage is a 64×64 two-dimensional image that forms the R channel of the final RGB image.

### 2.1.2. Channel G

To extract latent periodic structures in the vibration signal, this study employs an advanced cepstrum-based representation, the Advanced Cepstrum Feature Map. The primary goal of this channel is to transform raw frequency information into a structured, rich visual representation. The cepstrum can reveal information about the periodicity of harmonics, modulations, and repetitive structures present in the signal [25].

First, the DC component is removed from the one-dimensional signal x(n) to eliminate the bias and constant-amplitude variations:

$$x_c(n) = x(n) - \mu_x \tag{2}$$

$$\mu_x = \frac{1}{N}\sum_{n=0}^{N-1} x(n) \tag{3}$$

where $\mu_x$ is the signal mean. Removing the DC component reduces unnecessary effects caused by the offset. After that, the Hann window is used to reduce spectral leakage. It smooths signal edges and reduces spectral leakage in the Fourier transform [26]. The Hann window function is defined as follows:

$$\omega(n) = 0.5(1 - \cos(2\pi n/N)) \tag{4}$$

As a result, the windowed signal will be equal to:

$$x_\omega(n) = x_c(n) \times \omega(n) \tag{5}$$

Then the Fast Fourier Transform (FFT) is applied to the signal:

$$X(k) = \sum_{n=0}^{N-1} x_\omega(n) e^{-j2\pi kn/N} \tag{6}$$

where $X(k)$ is the frequency representation of the signal. Next, the logarithm of the absolute magnitude of the frequency spectrum is extracted:

$$L(k) = \log(|X(k)| + \varepsilon) \tag{7}$$

where $\varepsilon$ is a very small value to prevent an infinite result when taking the logarithm of zero. Using the spectral logarithm compresses the dynamic range and enhances weaker frequency components. Then, an inverse Fourier transform is applied to extract the signal's real cepstrum:

$$C(q) = \mathcal{F}^{-1} L(k) \tag{8}$$

where $q$ denotes Quefrency, a specific quantity in cepstral analysis that represents the repetition rate of spectral structures. Then, the cepstrum is divided into 64 windows, and 14 statistical and structural features are extracted for each window. These features include Mean, Standard Deviation, RMS, Skewness, Kurtosis, Peak Value, Normalized Peak Position, Energy, Entropy, Zero Crossing Rate, Linear Slope, Crest Factor, Dominant-to-Mean Cepstral Ratio, and Variance of Derivative. Then a 64×14 matrix is formed and, using interpolation, is converted into a 64×64 matrix. Finally, the resulting matrix is mapped to the [0, 255] range using Min-Max Normalization and stored as a grayscale image. This image represents the frequency-domain channel of the proposed system. In the final image, this channel is used as the green channel.

#### 2.1.3. Channel B

To simultaneously extract temporal and frequency information from the vibration signal, this study employs a hybrid structure called the Adaptive Time-Frequency Map (ATFM). The primary objective of designing this channel is to create a stable, rich two-dimensional representation of the signal's local variations. In this method, the TF channel is constructed from three main stages: Envelope Extraction, Continuous Wavelet Transform (CWT), and Energy Normalization.

First, the DC component is removed from the signal. Next, to better reveal the pulse patterns and modulations caused by the fault, the signal's envelope is extracted [27]. For this purpose, the Hilbert transform is used. First, the analytic signal is computed as follows:

$$z(t) = x_c(t) + j\hat{x}_c(t) \tag{9}$$

where $\hat{x}_c(t)$ is the Hilbert transform of the signal. Then the signal's envelope is obtained from the absolute value of the analytic signal:

$$e(t) = |z(t)| = \sqrt{x_c^2(t) + \hat{x}_c^2(t)} \tag{10}$$

After extracting the envelope, time-frequency analysis is performed using the CWT. The main idea of it is to compare the signal with different versions of a mother wavelet. This wavelet is both time-shifted and stretched or compressed across various scales. The main CWT equation is defined as follows:

$$W(a,b) = \frac{1}{\sqrt{a}}\int_{-\infty}^{\infty} x(t)\psi^*((t-b)/a)dt \tag{11}$$

where $\psi$ is the mother wavelet, $a$ is the scaling factor, and $b$ is the translation factor (time shift). In this equation, the $b$ factor determines at which time position the wavelet is located, and the $a$ factor determines how much the wavelet is stretched or compressed. The output of CWT produces a two-dimensional matrix with dimensions 64×4096, where each row corresponds to a specific scale and each column corresponds to a temporal position in the signal. After computing the wavelet coefficients, the time-frequency power is extracted using the squared absolute values of the CWT coefficients.

$$P(a,b) = |W(a,b)|^2 \tag{12}$$

This power matrix shows the signal's energy intensity across different scales and time points. However, because the energy of some scales may be much larger than that of others, a large portion of the weaker information would be lost without normalization. Therefore, energy normalization is applied:

$$P_n(a,b) = P(a,b)/\left(\sum\nolimits_b P(a,b) + \varepsilon\right) \tag{13}$$

where $\varepsilon$ is a very small value to prevent division by zero. Then, Log Compression is used to reduce the sharp discrepancy between large and small values:

$$P_{log}(a,b) = \log(1 + P_n(a,b)) \tag{14}$$

Finally, the resulting time-frequency matrix is interpolated to a fixed 64×64 size, then normalized and mapped to the luminance range [0,255] to produce a standard grayscale image. This image is used as the blue channel in the final RGB structure.

### 2.2. Masking Images

In this study, the problem of reconstructing corrupted data has been chosen as an auxiliary task. To this end, part of the input image is masked. This process ensures that the model is not only dependent on surface-level features but also learns structural connections, spatial patterns, and dependencies between different signal domains. In fact, when part of the data is hidden, the model is forced to estimate the overall structure using surrounding information. In addition, the input images have three channels, so when part of an image is masked, the model must learn the interactions among these three domains to reconstruct the removed regions. In this study, three different masking strategies are investigated: Patch Masking, Donut Masking, and Hybrid Masking.

#### 2.2.1. Patch Masking

In this method, the input image is divided into equally sized square blocks (patches), and some of these patches are randomly removed. It causes the model to learn spatial dependencies and semantic-level interactions from the context and to place greater importance on global features [28]. Examples of this masking method are shown in Figure 3. If the input image is assumed to have dimensions H×W and each patch size is P×P, the number of divisible patches is defined as follows:

$$N_H = H/P \tag{15}$$

$$N_W = W/P \tag{16}$$

And the total number of patches will be equal to:

$$N = N_H \times N_W \tag{17}$$

In this study, P = 32 was considered. Then, a percentage of the patches is removed according to the masking ratio. If the masking ratio is r, the number of masked patches is obtained from the following equation:

$$N_M = r \times N \tag{18}$$

In this model, r is set to 0.35, which means that approximately 35 percent of all image patches are removed. The important point about this number is that increasing it (r = 0.6) improves denoiser performance, while decreasing it (r = 0.15) improves classifier performance. Therefore, the optimal r lies between these two values. Then, a mask matrix is generated, with each element indicating whether the corresponding patch has been removed. This mask can be defined as follows:

$$M_{Block}(i,j) = \begin{cases} 1 & If\ the\ patch\ has\ been\ removed \\ 0 & Otherwise \end{cases} \tag{19}$$

Then the masked image is created from the following equation:

$$I_{masked} = I \odot \big(1 - M(i,j)\big) \tag{20}$$

where $I$ is the original image and $\odot$ denotes element-wise multiplication. In fact, the selected patches are set to zero, and the network uses only the remaining information for reconstruction.

The patch masking method, by removing contiguous regions of the image, has greater capacity to learn semantic and contextual features. This leads to deeper feature extraction, the learning of long-range dependencies, and reduced model dependence on low-level details. Moreover, since sensor faults, severe noise, and data loss in vibration data typically occur in contiguous regions, patch masking introduces a realistic form of data corruption. Despite these advantages, patch masking also has limitations: the square, regular structure of the patches may not perfectly match real fault patterns in industrial data, and, in some cases, removing small critical areas can lead to the loss of important information and a decrease in accuracy [29].

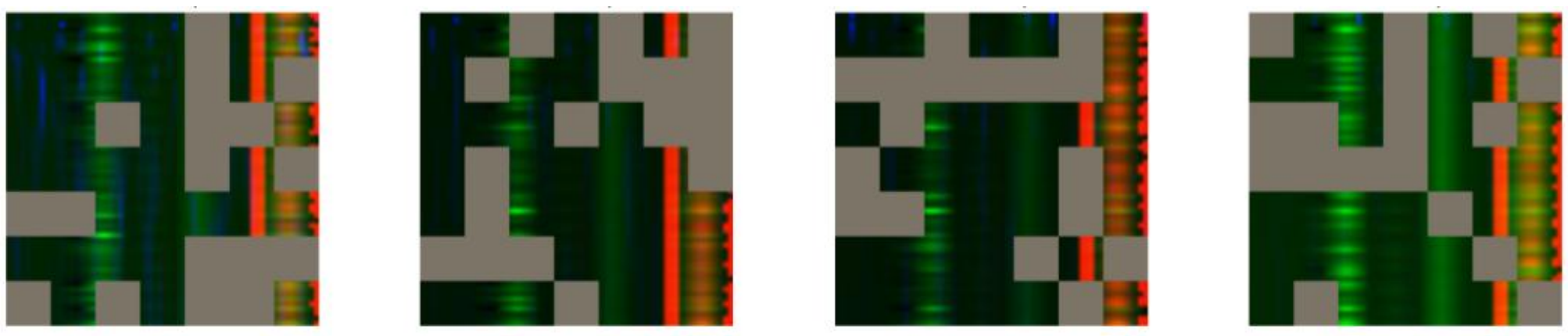

*Figure 3: Examples of Patch-Based Masking Applied to Input Images.*

### 2.2.2. Donut Masking

Another masking method is donut masking, in which the removed regions are defined as circular or ring-shaped. The main goal of this method is to generate more natural erosion patterns that more closely reflect real-world industrial data conditions. Also, in this method, only the central region of the donut is removed, and the surrounding ring remains as the context. Therefore, the model must reconstruct the lost information solely using nearby neighbors, and its primary focus is on local features. This is a very important difference from patch masking. In patch masking, the model may use regions from very distant parts of the image for reconstruction. In contrast, in donut masking, the model is effectively forced to learn local dependencies and nearby spatial continuity. Examples of this masking method are shown in Figure 4.

Initially, for each image, several random centers are selected:

$$\left(c_x^k, c_y^k\right), k = 1,2, \dots, K \tag{21}$$

where K is equal to the number of donuts. Then, for each center, the Euclidean distance of all pixels to the center is calculated:

$$D(x, y) = \sqrt{(x - c_x)^2 + \left(y - c_y\right)^2} \tag{22}$$

Then the outer radius is defined:

$$R_{outer} = S/2 \tag{23}$$

where $S$ is the size of the mask area, set to 32. Then the inner radius is calculated as a fraction of the outer radius:

$$R_{inner} = \alpha R_{outer} \tag{24}$$

where, $\alpha = 0.5$ has been considered. After that, the pixels located within the inner radius are masked:

$$M_{Donut}(x, y) = \begin{cases} 1 & D(x, y) \le R_{inner} \\ 0 & Otherwise \end{cases} \tag{25}$$

where $R_i$ is the inner radius. With this method, only the donut's center is removed, and the masked image is generated using equation 20.

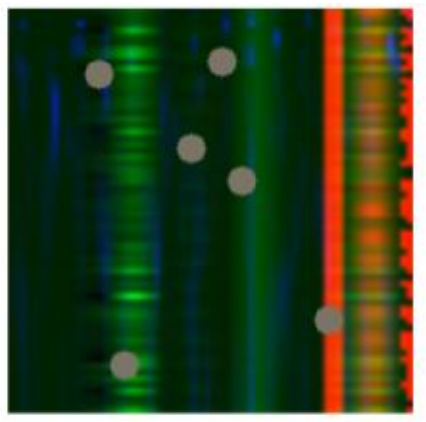 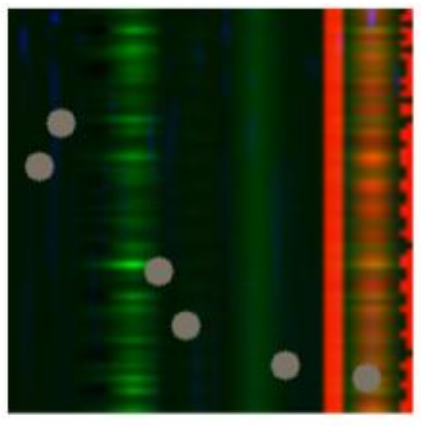 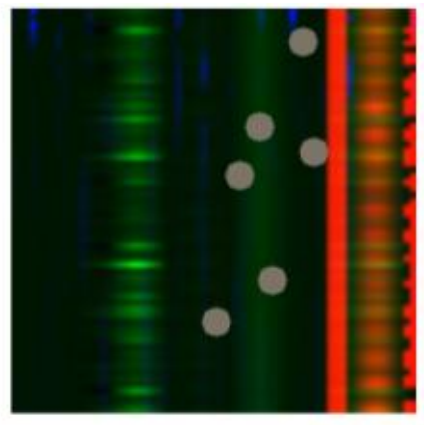 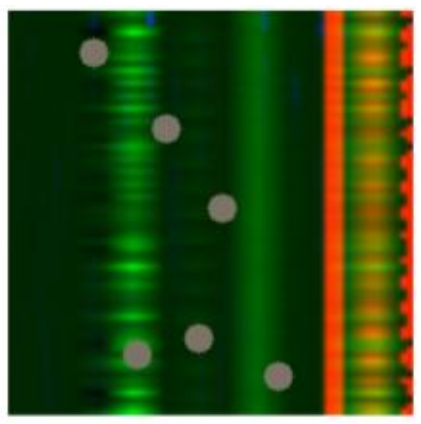

*Figure 4: Examples of the Proposed Donut Masking Strategy.*

### 2.2.3. Hybrid Masking

As mentioned, the patch masking method primarily learns global structures, high-level features, and long-range dependencies. In contrast, the donut masking method places greater emphasis on local details, fine-grained patterns, ridges, low-level features, and continuous structures. Therefore, using only one of these two methods can lead to unbalanced feature learning, causing the network to either become overly dependent on the global context or focus solely on local details. For this reason, this study designs a hybrid masking method that simultaneously leverages the advantages of both the patch and donut strategies. The core idea of this approach is that during training, the network is randomly exposed to both types of data corruption, enabling it to learn both global dependencies and local features simultaneously. The overall process is defined as follows:

$$M(x) = \begin{cases} M_{Donut}(x) & p < 0.5 \\ M_{Block}(x) & p \geq 0.5 \end{cases} \tag{26}$$

where $p$ is a uniform random number in the interval [0,1]. The combination of these two strategies in the hybrid method has enabled the model to engage in two distinct types of learning simultaneously. On the one hand, donut masking forces the network to learn local details, fine frequency patterns, continuous lines, and sensitive structures. On the other hand, patch masking compels the model to grasp global dependencies and overall image structures. As a result, the extracted features will be richer, and the model will not be dependent on just one type of context. Examples of this method are shown in Figure 5.

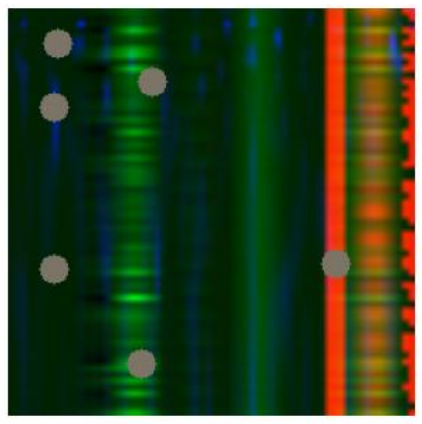 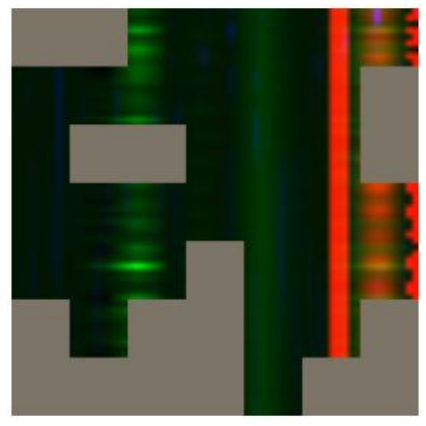 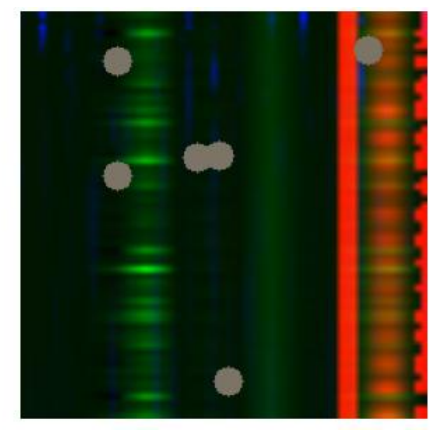 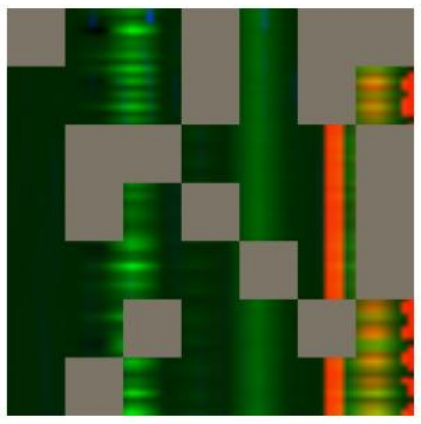

*Figure 5: Examples of the Proposed Hybrid Masking Strategy.*

## 2.3. Encoder

in this study, a multi-task encoder based on transfer learning was designed to extract deep and robust features from RGB images. Within the proposed XMDCA-TL framework, the backbone of this section is based on the ConvNeXtV2 architecture and is combined with Multi-Domain Channel Attention (MDCA) modules and a denoising module. The goal is not only to increase classification accuracy but also to generate richer feature representations of the signal's physical and domain aspects. The network's input consists of masked RGB images composed of three domains. For this reason, the encoder is designed to directly model the dependencies and interactions among these three channels.

Additionally, the fault diagnosis and self-supervised denoising processes are performed simultaneously during training. This feature leads to more robust representations in the face of noise, incomplete data, and domain shifts, enhancing the model's generalization capability across various operational conditions. An overall diagram of the encoder's structure is presented in Figure 6.

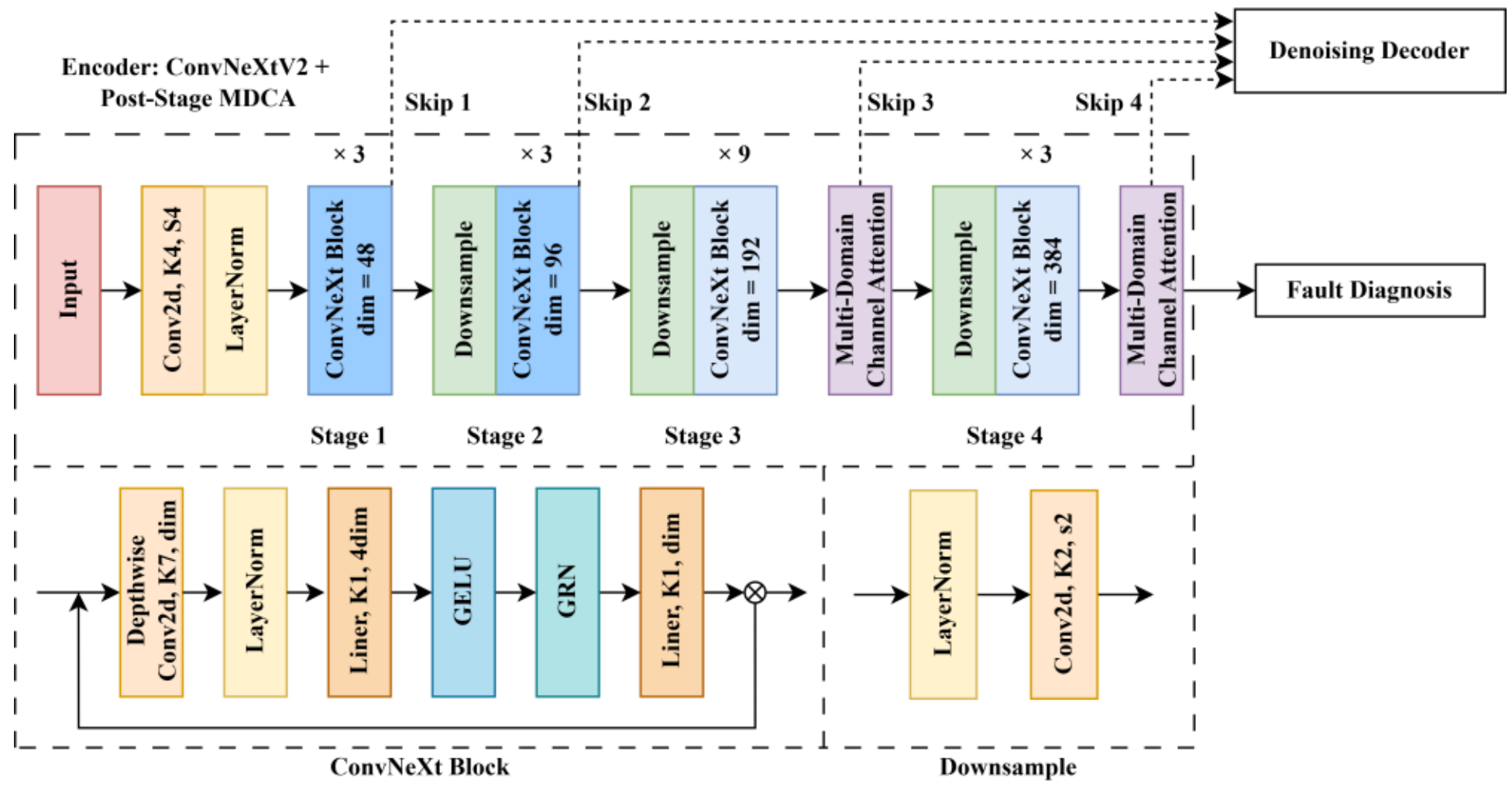

*Figure 6: Architecture of the Proposed XMDCA-TL Encoder.*

The ConvNeXtV2 architecture consists of several sequential stages, each stage containing multiple advanced convolution blocks. In each block, depthwise convolution with a large kernel is first performed to extract spatial dependencies with a wide receptive field. Then, layer normalization and two pointwise linear layers with the GELU activation function are applied. In the ConvNeXtV2 version, a Global Response Normalization (GRN) module has also been added, which helps balance channel responses and prevent the dominance of certain feature maps [16]. The general structure of each block can be expressed as follows:

$$X_{out} = X + F(X) \tag{27}$$

where $X$ is the block input and $F(X)$ is the feature extraction function. The presence of a residual connection provides gradient stability, prevents vanishing gradients, and facilitates the training of deep networks [30]. In ConvNeXtV2, the depthwise convolution operation is defined as follows:

$$Y_c = K_c * X_c \tag{28}$$

where $X_c$ is the input channel, $K_c$ is the corresponding kernel, and $*$ indicates convolution. Unlike standard convolution, which processes all channels simultaneously, depthwise convolution filters each channel independently. This leads to a significant reduction in the number of parameters and computational cost [31]. This architecture also uses the Gaussian Error Linear Unit (GELU) activation function, which has a smoother and more continuous behavior than ReLU:

$$GELU(x) = x\Phi(x) \tag{29}$$

where $\Phi(x)$ is the cumulative Gaussian distribution function. Using GELU improves gradient flow and enables more stable learning of complex features. One of the most important components of ConvNeXtV2 is the GRN module, which works as follows:

$$G_x = \|x\|_2 \tag{30}$$

$$N_x = (G_x)/(mean(G_x) + \varepsilon) \tag{31}$$

$$Y = \gamma(xN_x) + \beta + x \tag{32}$$

In these relations, $G_x$ is the overall energy of the feature map, and $N_x$ is its normalized version. The parameters $\gamma$ and $\beta$ are learnable. The GRN enables the network to better regulate dependencies among channels and produce a more balanced feature representation.

The ConvNeXtV2 model uses initial weights pre-trained on the ImageNet dataset. Using pretrained weights ensures that the network has an initial understanding of visual structures, edges, frequency patterns, and high-level features right from the start of training. This is especially important in scenarios where data volume is limited or there is a domain gap between the source and target datasets. In addition, the backbone is not trained fully from scratch; instead, the freeze/unfreeze strategy is used. In the initial stages, most of the backbone layers are frozen, and only the network's final stages and the layers related to classification and attention are trained. This preserves the initial knowledge gained from pretraining and prevents overfitting. The backbone's learning rate is set lower than that of the other components to avoid sudden degradation of the pretrained weights.

In the next step, the MDCA mechanism was designed as a specialized attention for vibration data to directly model the cross-dependencies among the three physical domains. This mechanism is shown in Figure 7. The MDCA structure is designed based on the cross-attention mechanism. Unlike self-attention, which extracts query, key, and value from a single source, cross-attention allows features from one domain to receive information from another. This process allows each domain to leverage complementary information available in the other domains.

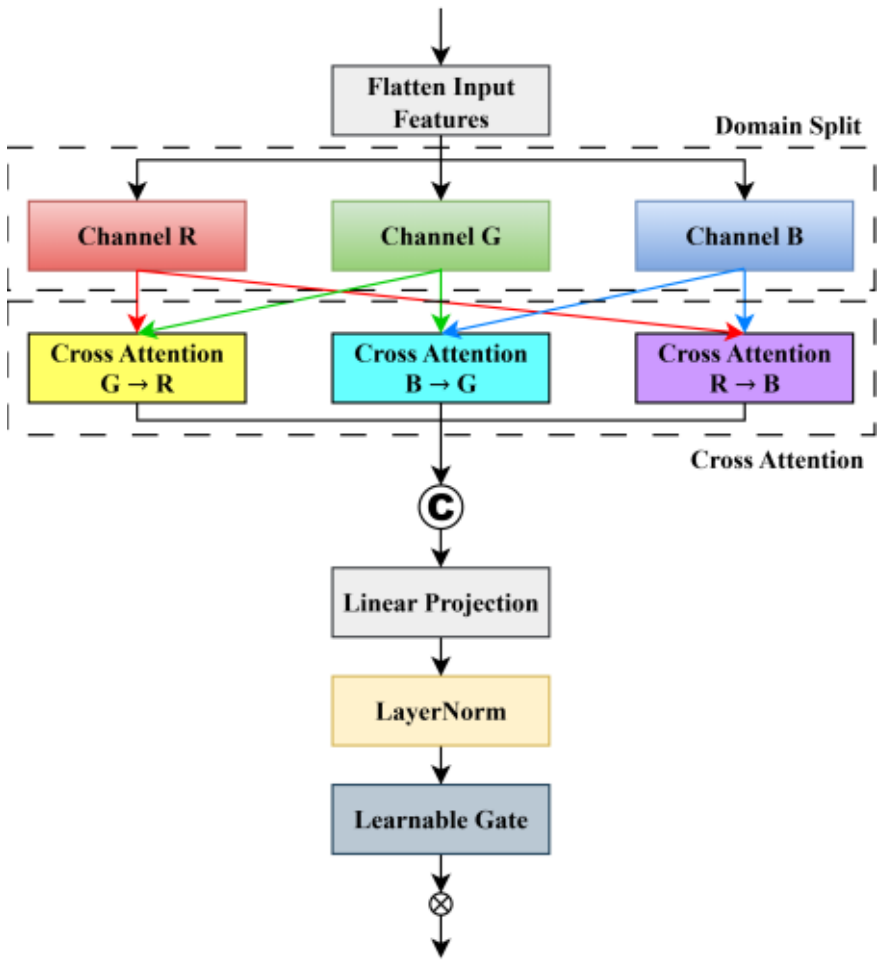

*Figure 7: Structure of the Proposed Multi-Domain Channel Attention Mechanism.*

First, the output feature map of the backbone is divided into three channel groups corresponding to the three domains:

$$X = [X_t, X_f, X_{tf}] \tag{33}$$

where $X_t$ represents the features in the time domain, $X_f$ represents the features in the frequency domain, and $X_{tf}$ represents the features in the time-frequency domain. After splitting the feature map, three independent cross-attention paths are defined as shown in Figure 8.

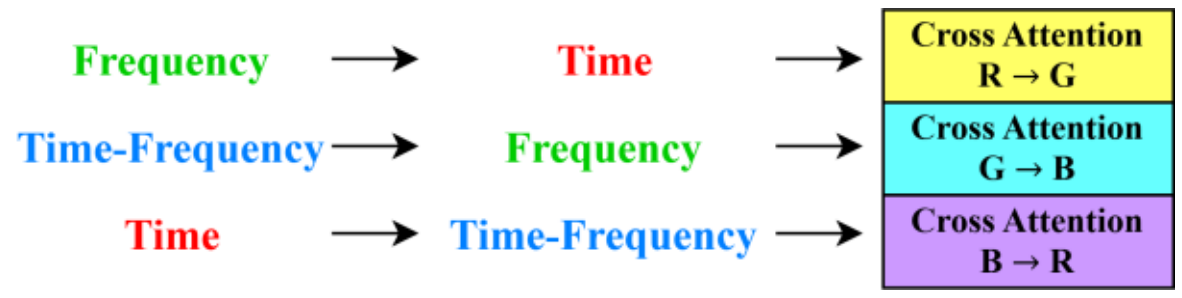

*Figure 8: Cross-Attention paths.*

As a result, each information domain receives its complementary information from another domain. For example, temporal features can learn from the frequency patterns associated with fault frequencies, and time-frequency domain features can also leverage the signal's temporal variations. This process creates a physical dependency among the different signal representations. In the attention mechanism, the Query, Key, and Value matrices are first generated [32]:

$$Q = XW_Q \tag{34}$$

$$K = XW_K \tag{35}$$

$$V = XW_V \tag{36}$$

Then the attention weights are calculated as follows:

$$Attention(Q, K, V) = Softmax\left(QK^T/\sqrt{d}\right)V \tag{37}$$

where $d$ denotes the feature dimension. Using the $\sqrt{d}$ coefficient stabilizes the gradients and prevents attention values from growing excessively. Then, the outputs of three different attentions are merged and returned to the original feature space via a projection layer:

$$O = W_o\left[O_t, O_f, O_{tf}\right] \tag{38}$$

Finally, the attention output, after passing through a layer normalization layer, is added to the original feature map as a residual connection:

$$\hat{O} = LayerNorm(O) \tag{39}$$

$$Y = X + \alpha\hat{O} \tag{40}$$

where $\alpha$ is a learnable gate. This parameter is initialized to zero at the start of training, allowing the model to use the pretrained weights of ConvNeXtV2 without interference. Then, during training, the model gradually learns how much attention information to incorporate into the final representation. The use of this gated residual mechanism is considered a key component of the MDCA design, as it prevents degradation of pretrained features during the early stages of fine-tuning. Also, to prevent severe instability, the gate value is limited by the hyperbolic tangent function:

$$Y = X + \tanh(\alpha)\,\hat{O} \tag{41}$$

One of the most important aspects of this study's design is the placement of the MDCA within the ConvNeXtV2 architecture. Initially, the goal was to embed the MDCA directly into the ConvNeXtV2 blocks – applying it after the GRN layer and before the final projection. However, implementing this structure led to a dramatic increase in the number of parameters. Investigations showed that embedding the MDCA in all ConvNeXtV2 blocks adds approximately 100 million additional parameters to the model. This sharp increase in parameters not only led to higher GPU memory usage and longer training times but also significantly increased the likelihood of overfitting. For this reason, in the final design, it was decided to place the MDCA outside the inner backbone blocks and apply it only after the outputs of the backbone's deeper stages, specifically after stages 3 and 4 of the network. This decision provided several important advantages. First, the pretrained weights of ConvNeXtV2 were almost completely preserved, and the core backbone structure remained intact. Second, the number of parameters was dramatically reduced, making the model much more computationally lightweight. Third, applying attention in deeper stages allows MDCA to operate on feature maps that contain stronger semantic information, thereby more effectively extracting cross-domain dependencies.

Moreover, the attention mechanism enables the model to adaptively focus on the more important channels. In fault diagnosis, many features contain unnecessary or noisy information, and not all channels are equally important. MDCA, by learning the attention weights, amplifies channels that carry more relevant information about fault patterns and attenuates those that carry less relevant information.

In this architecture, transfer learning is performed in multiple stages. First, the network is trained on data from the source domain. Then, in the cross-domain adaptation stage, the model is evaluated on the target domain, and finally, fine-tuning is performed on the target data. This process enables the encoder to produce representations that, in addition to high discriminative power, are sufficiently stable against domain shifts. In other words, the model is not limited to learning a simple mapping; rather, it attempts to extract features that are transferable across different domains.

Moreover, the output of each stage, after feature extraction by the backbone and the application of the MDCA module via a skip connection, is stored as a skip feature. Therefore, the decoder can use feature maps at every level of reconstruction that contain both local convolutional information and cross-domain dependencies. This design ensures that important information about the vibration signals is not lost during encoding and decoding.

### 2.4. Denoising Decoder

In the following, the denoising decoder, which is based on the UNet architecture, is examined. Its primary task is to reconstruct input images and estimate missing information. The overall flowchart of the decoder network is shown in Figure 9.

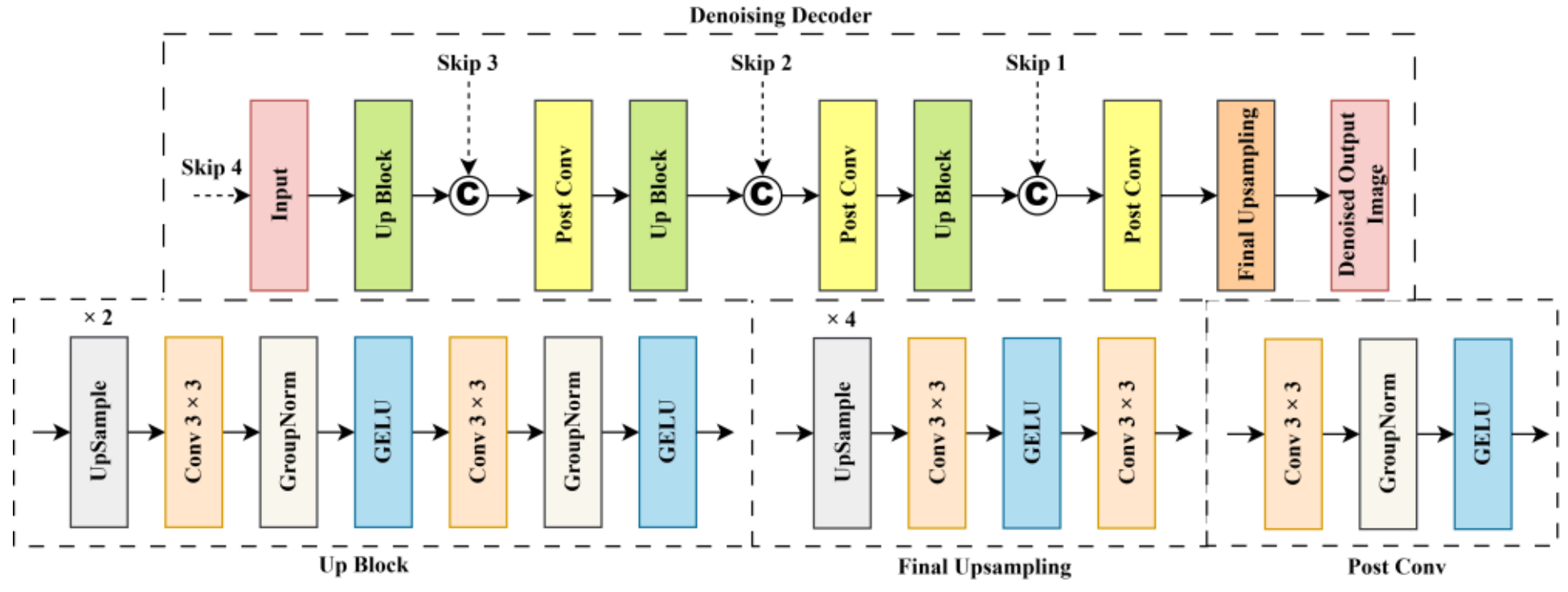

*Figure 9: Structure of the UNet-Based Denoising Decoder.*

If the original image is denoted by X, the masked image by $\tilde{X}$, and the decoder output by $\hat{X}$, the decoder's objective is to minimize the difference between the reconstructed image and the original image:

$$\hat{X} = D\left(E(\tilde{X})\right) \tag{42}$$

where $E$ denotes the encoder, and $D$ denotes the decoder. In the proposed architecture, the Decoder is directly connected to the multiscale features extracted by the encoder. To achieve this, a skip connection structure similar to that of U-Net is used. It enables the simultaneous use of low- and high-level information during reconstruction. From a mathematical perspective, if the output of Stage $i$ in the encoder is denoted by $F_i$, the set of features extracted by the encoder can be represented as follows:

$$F = \{F_1, F_2, F_3, F_4\} \tag{43}$$

where:

$$F_i \in R^{C_i \times H_i \times W_i} \tag{44}$$

and sequentially includes the extracted feature maps at different network levels. The decoder begins reconstruction from the deepest feature representation, namely $F_4$.

The decoder in this architecture consists of three main components: the Up Block, the Post-Concatenation Convolution Block (Post-Conv), and the Final Upsampling Block. The task of the Up Block (upsampling block) is to gradually increase the resolution of the feature maps extracted by the encoder. As in the encoder, the spatial dimensions of the image are reduced during feature extraction; in the decoder, they must be restored to their original size to enable image reconstruction. This upsampling operation is performed using Bilinear Interpolation:

$$H_{out} = 2H_{in}, W_{out} = 2W_{in} \tag{45}$$

where the spatial dimensions of the feature map are doubled at each stage. Unlike Transposed Convolution-based methods, Bilinear Upsampling offers several advantages. Its most significant benefit is the prevention of checkerboard artifacts, a well-known issue in image reconstruction [33]. Additionally, this method has no learnable parameters, which reduces the model's parameter count and the likelihood of overfitting. After upsampling, two consecutive convolution layers with a 3×3 kernel are applied to the feature map:

$$F_{mid} = \sigma\left(GN\left(Conv_{3\times3}(Up(F_{in}))\right)\right) \tag{46}$$

$$F_{out} = \sigma\left(GN(Conv_{3\times3}(F_{mid}))\right) \tag{47}$$

where $Up$ denotes the upsampling operation, $Conv$ denotes convolution, $GN$ denotes GroupNorm normalization, and $\sigma$ denotes the GELU activation function.

In all decoder blocks, after each convolution layer, Group Normalization (GN) is followed by GELU activation. This architecture ensures that the feature maps produced during image reconstruction have greater statistical stability while preserving the ability to model complex nonlinear interactions. In GN, the channels of a feature map are divided into G groups, and then for each group, the mean and variance are calculated as follows:

$$\mu_g = (1/m)\sum_k x_k \tag{48}$$

$$\sigma_g^2 = (1/m)\sum_k (x_k - \mu_g)^2 \quad (49)$$

where $m$ is the total number of elements in that group. Then the normalization operation is performed as follows:

$$\hat{x}_i = (x_i - \mu_g)/\left(\sqrt{\sigma_g^2 + \varepsilon}\right) \quad (50)$$

Finally, the learnable scale and bias parameters are applied:

$$y_i = \gamma\hat{x}_i + \beta \quad (51)$$

where $\gamma$ and $\beta$ are trainable parameters. After normalization, the GELU activation function is used. After each upsampling stage, the generated feature map is merged with the corresponding encoder feature map via a skip connection. This merging is performed as channel-wise concatenation:

$$F_{cat} = Concat(F_{up}, F_{skip}) \quad (52)$$

As a result, the number of channels nearly doubles. If this feature map is sent directly to the next stage, several problems arise: an excessive increase in the number of channels, statistical discrepancies between the encoder and decoder features, increased noise and redundant information, and higher computational cost. For this reason, after concatenation, a Post-Conv block is used, which includes a 3×3 convolution, GroupNorm, and GELU.

After the data passes through all stages, a final upsampling stage is applied to restore the image dimensions to the input resolution. This operation is performed with a factor of four:

$$H_{final} = 4H_{decoder}, W_{final} = 4W_{decoder} \quad (53)$$

After this stage, two consecutive convolution layers are applied. The first layer reduces the number of channels to 32 and serves as the final feature refinement. The second layer also produces the output RGB image with three channels:

$$F_{32} = GELU(Conv_{3\times3}(F)) \quad (54)$$

$$I_{out} = Conv_{3\times3}(F_{32}) \quad (55)$$

where $I_{out}$ is the final reconstructed image.

### 2.5. Fault Diagnosis

In the proposed architecture, the primary fault diagnosis task is entrusted to a lightweight, purpose-built classification head. It follows immediately after the ConvNeXtV2-based encoder and the MDCA modules. First, the final feature map is normalized using layer normalization to stabilize the feature distribution and reduce the impact of domain shifts. Then, global average pooling is applied across all spatial locations. This operation compresses the spatial information in the feature map into a compact vector. This vector contains aggregated information from the entire RGB image and represents the most important features related to the type of corruption. In the next step, the feature vector is fed into a fully connected layer to compute the sample's probability of belonging to each class. Finally, the probability that the sample belongs to each damage class is computed using the Softmax function.

Since the network's backbone is built on a pre-trained ConvNeXtV2 model on ImageNet, not all network parameters can be updated at the start of training. To prevent the loss of pre-learned knowledge and reduce the number of trainable parameters, a partial fine-tuning strategy is employed. In this strategy, all backbone parameters are first frozen, and only the final two Stages of the network remain unfrozen. This approach ensures that the initial layers, which extract general features, remain unchanged, while the deeper layers are adapted to learn domain-specific features. The Adam algorithm is used for network optimization. To prevent drastic changes to the pre-trained features, the learning rate for the backbone layers is set to 1/10 of that for the other parts of the network.

The training pipeline is designed based on the gradual transfer of knowledge from the source domain to the target domain. Initially, the network is trained on source-domain data. The main objective of this stage is to extract general and robust feature representations from vibration signals. Then, its performance is evaluated on the target domain without any adaptation to measure the domain gap. In the final stage, the network is fine-tuned on target-domain data, and all parameters are freed to adapt. This process allows the extracted features to align with the target domain's data distribution while preserving prior knowledge. There is a discrepancy between the source and target data, known as domain shift. Changes in operating conditions, sensor differences, variations in rotational speed, noise levels, loading conditions, and other environmental factors cause it. Therefore, it is necessary to expose the model to target domain data and fine-tune its parameters to align its feature representations with the new distribution.

Since the proposed architecture is designed as a multitask system, the network training process simultaneously pursues two main objectives. To achieve both objectives simultaneously, a combined loss function was defined that includes the classification error and the image reconstruction error. The overall model loss function is defined as follows:

$$L_{total} = \lambda_{cls} L_{cls} + \lambda_{den} L_{den} \tag{56}$$

where $L_{cls}$ is the classification error, $L_{den}$ is the image reconstruction error, $\lambda_{cls}$ is the importance coefficient for the classification task, and $\lambda_{den}$ is the importance coefficient for the reconstruction task. According to the model settings, the coefficients were set to $\lambda_{cls} = 1$ and $\lambda_{den} = 0.2$. Hence, the network's primary task remains fault diagnosis, while the reconstruction task serves as an auxiliary mechanism for learning more robust features.

The Cross-Entropy Loss function was used to train the fault diagnosis branch. This function is one of the most common optimization metrics in multiclass classification problems, measuring the discrepancy between the network's predicted probability distribution and the actual labels. The formula for Cross-Entropy is as follows:

$$L_{cls} = (-1/N) \sum_{i=1}^{N} \sum_{c=1}^{C} y_{ic} \log(\hat{y}_{ic}) \tag{57}$$

where $N$ is the number of samples, $C$ is the number of classes, $y_{ic}$ is the actual label, $\hat{y}_{ic}$ is the predicted probability by Softmax.

The Mean Squared Error (MSE) function is used to measure the quality of the reconstruction. In this study, two types of reconstruction errors are calculated separately. The first error is the error in the masked regions, which is the most important part of the reconstruction process:

$$L_{masked} = (1/N_m) \sum_{i \in \Omega_m} (x_i - \hat{x}_i)^2 \tag{58}$$

where $\Omega_m$ is the set of masked pixels, and $N_m$ is the number of masked pixels. This loss forces the network to reconstruct the missing information solely from the image's overall structure and knowledge extracted from deep features. In addition to the removed regions, a reconstruction error is also calculated for the intact regions of the image to ensure that the overall image reconstruction remains structurally stable.

$$L_{unmasked} = (1/N_u) \sum_{i \in \Omega_u} (x_i - \hat{x}_i)^2 \quad (59)$$

where $\Omega_u$ is the set of pixels without a mask and $N_u$ is the number of these pixels. Since reconstructing the removed regions is more important than reconstructing the intact regions, different coefficients were assigned to the two regions. The reconstruction cost function is defined as follows:

$$L_{den} = \lambda_{masked} L_{masked} + \lambda_{unmasked} L_{unmasked} \quad (60)$$

where in the current implementation $\lambda_{masked} = 1$ and $\lambda_{unmasked} = 0.3$. Thus, the final cost function used in the network's training is as follows:

$$L_{total} = L_{cls} + 0.2(L_{masked} + 0.3 L_{unmasked}) \quad (61)$$

### 2.6. XAI

To clarify the decision-making process, an Explainable Artificial Intelligence (XAI) framework has been designed, with its overall structure and execution steps shown in Figure 10. This methodology, aimed at addressing the model's black-box nature, consists of four main phases. The implementation of this hierarchical process enables the physical analysis of the model's behavior and the rigorous validation of the knowledge transfer approach.

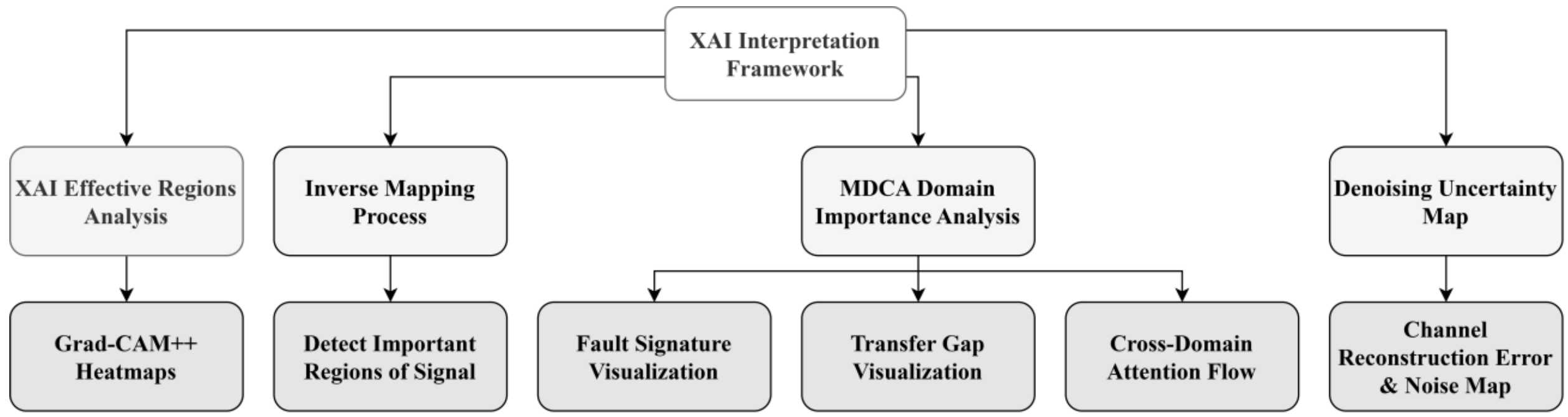

*Figure 10: Flowchart of the Proposed Explainable Artificial Intelligence Framework.*

#### 2.6.1. Analysis of Effective Regions in Model Decision-Making

The purpose of this section is to identify the effective regions of the image for fault diagnosis. Although deep learning models can achieve high accuracy, their black-box nature makes it unclear which part of the data played the primary role in the final decision. For this reason, the XAI method based on Grad-CAM++ was used to determine the importance of different regions of the RGB images derived from the vibration signals [34].

The Grad-CAM++ method is an extension of Grad-CAM that, by leveraging higher-order gradients, can pinpoint the regions influencing the model's prediction with greater accuracy. By calculating the importance of each feature map with respect to the output class, this method generates a heatmap that shows which parts of the image the model paid more attention to during its decision-making.

In the Grad-CAM++ method, the output gradient for the target class $y^c$ is first calculated with respect to the feature maps of the selected layer. The importance weight of each feature map $A^k$ for class $c$ is calculated as follows [35]:

$$w_k^c = \sum_i \sum_j a_{ij}^{kc} ReLU\left(\partial y^c / \partial A_{ij}^k\right) \tag{62}$$

where $A_{ij}^k$ is the feature mapping value at position $(i, j)$ and channel $k$, and $a_{ij}^{kc}$ is a coefficient calculated using the second- and third-order derivatives of the network output:

$$a_{ij}^{kc} = \left(\partial^2 y^c / \left(\partial A_{ij}^k\right)^2\right) \Big/ \left(2\left(\partial^2 y^c / \left(\partial A_{ij}^k\right)^2\right) + \sum_{a,b} A_{ab}^k \left(\partial^3 y^c / \left(\partial A_{ij}^k\right)^3\right)\right) \tag{63}$$

After calculating the importance weights, the final Grad-CAM++ map is obtained as follows:

$$L_{GradCAM++}^c = ReLU\left(\sum_k w_k^c A^k\right) \tag{64}$$

where $L_{GradCAM++}^c$ is the class-specific importance map for class $c$. Applying the ReLU function ensures that only regions that positively impact the model's decision-making are retained. As a result, the heatmap shows which parts of the image played the greatest role in the network's final prediction. To analyze the model's behavior, Grad-CAM++ maps were extracted from the output of the last MDCA module. This layer was chosen because it lies in the deepest part of the feature extraction and is closest to the final decision-making process.

#### 2.6.2. Inverse Mapping of XAI Results from the Image Space to the Vibration Signal Space

Since the system's primary input is the vibration signal and the RGB images are merely representations extracted from it, analyzing the results solely in the image domain cannot provide a complete understanding of the system's physical behavior. For this reason, at this stage, a back-projection process was designed to back-project the important regions identified in the image back onto the original vibration signal and determine which temporal segments of the signal played the greatest role in the model's decision-making.

To assess the importance of each time point, the average heatmap value along the image's vertical axis was calculated. As a result, for each column $c$ of the image, the importance value $I(c)$ was obtained as follows:

$$I(c) = (1/M) \times \sum_{r=1}^{M} H(r, c) \tag{65}$$

where $H(r, c)$ is the heatmap value at row $r$ and column $c$, and $M$ is the number of image rows. Then the value of $I(c)$ was assigned to the corresponding time interval in the signal, thereby forming a time-importance vector. This vector shows the contribution of each segment of the vibration signal to the network's final decision. In addition to the overall mapping, the importance of features in different channels was also examined separately. In the time channel, the average importance of each row was used to evaluate the role of time windows, whereas in the time-frequency channel, because the horizontal axis of the image directly corresponds to time, the importance of each column was mapped directly to its corresponding temporal position. Finally, to reduce local fluctuations and achieve a more stable representation, the resulting importance vector was smoothed with a Gaussian filter.

#### 2.6.3. Analysis of the Importance of Domains in MDCA

The purpose of this section is to examine the relative contribution of each of the physical domains – time, frequency, and time-frequency. Analyzing the contribution of each channel can provide valuable insight into the behavior of the MDCA module and how it learns inter-domain interactions. To this end, the average energy of the MDCA output features was used as a metric. After performing the cross-attention process, the energy of each domain was calculated as the mean squared activation energy:

$$E_d = (1/N)\sum_{i=1}^{N} o_{d,i}^2 \tag{66}$$

where $o_d$ is the cross-attention output for domain $d$. This metric indicates how much effective information from each domain was utilized in feature extraction. Domains that require more activation energy contribute more to the representation of the network's final features.

In the first step, the relative importance of each domain for the entire dataset was calculated to determine which physical representation the model relies on most at a general level. In the second step, the importance of domains was extracted separately for each fault class. This analysis enables examining the dependence of each fault type on temporal, frequency, or time-frequency information and can reveal which physical patterns each fault type uses for diagnosis.

##### *2.6.3.1. Fault Signature Visualization*

The purpose of this section is to examine the dependence of each fault class on three information domains. To this end, the relative contribution of each domain was calculated separately for samples belonging to each class. Thus, the mean energy of the features produced by the cross-attention paths in the final MDCA module for each class was extracted and normalized. This analysis enables the identification of the dominant domain for diagnosing each fault type. It shows how much the model relies on temporal, frequency, or time-frequency information to identify different classes.

##### *2.6.3.2. Transfer Gap Visualization*

In this subsection, the changes in the pattern of multi-domain information usage before and after the knowledge transfer process were examined. The purpose of this analysis is to evaluate the effect of fine-tuning on the model's use of temporal, frequency, and time-frequency information in the target domain. To this end, the relative importance of each domain in the two models – before transfer (source model) and after adaptation to the target domain (adapted model) – was calculated and compared. The evaluation metric was defined using feature energy to assess the impact of knowledge transfer on the model's attention distribution across domains.

##### *2.6.3.3. Cross-Domain Attention Flow*

To investigate how the three domains interact in the MDCA module, the attention flow between domains was analyzed. The goal is to know how much each physical representation can leverage information from the others. Each domain updates its features using information extracted from another domain (Time ← Frequency, Frequency ← Time-Frequency, and Time-Frequency ← Time). To this end, the attention matrices generated in the final MDCA module were extracted, and the attention intensity for each path was calculated by averaging the attention coefficients across paths.

#### 2.6.4. Denoising Uncertainty Map

Since the decoder attempts to remove unnecessary and noisy components while preserving meaningful information related to the defect during training, the difference between the input image and the reconstructed image can serve as a metric for identifying areas of uncertainty or noise. Accordingly, to analyze the model's behavior, the absolute difference between the original image and the reconstructed output was calculated for each pixel. This difference indicates which parts of the data the model identified as less reliable and to which it applied more changes during reconstruction. To this end, the reconstruction error for each channel was calculated as the absolute difference between the input and output values. In general, a larger reconstruction error for a channel indicates that the domain contains more noisy components. Furthermore, uncertainty analysis was conducted independently for different fault classes to determine in which domain each fault type exhibits the greatest ambiguity or noise. This analysis shows which temporal, frequency, or time-frequency representations the model identifies as requiring greater denoising when confronted with different fault patterns.

## 3. Dataset Overview

### 3.1. Benchmark Dataset

In this study, to evaluate the performance of the proposed model, the standard CWRU (Case Western Reserve University) dataset is used [36]. The experimental device is shown in Figure 11. Vibration data were collected using accelerometers mounted on the motor with a sampling frequency of 12 kHz. They included four main bearing conditions: healthy, inner-race fault, outer-race fault, and ball fault. For each fault type, different severity levels with diameters of 7, 14, and 21 mil were artificially created using the Electrical Discharge Machining (EDM) method. As a result, there are 10 distinct bearing health conditions in this dataset. The data were also recorded under various operating conditions, including different rotational speeds and motor loads, allowing evaluation of the model's performance under diverse operational scenarios.

Although the CWRU dataset includes several operational load levels, this study selected only the data corresponding to the zero-horsepower load as the source domain. This choice was made to reduce the influence of variables arising from changes in operating conditions and to focus on analyzing fault characteristics. Furthermore, preliminary experiments showed that using all load conditions simultaneously does not significantly affect the model's final performance, but it increases training computational cost and complexity.

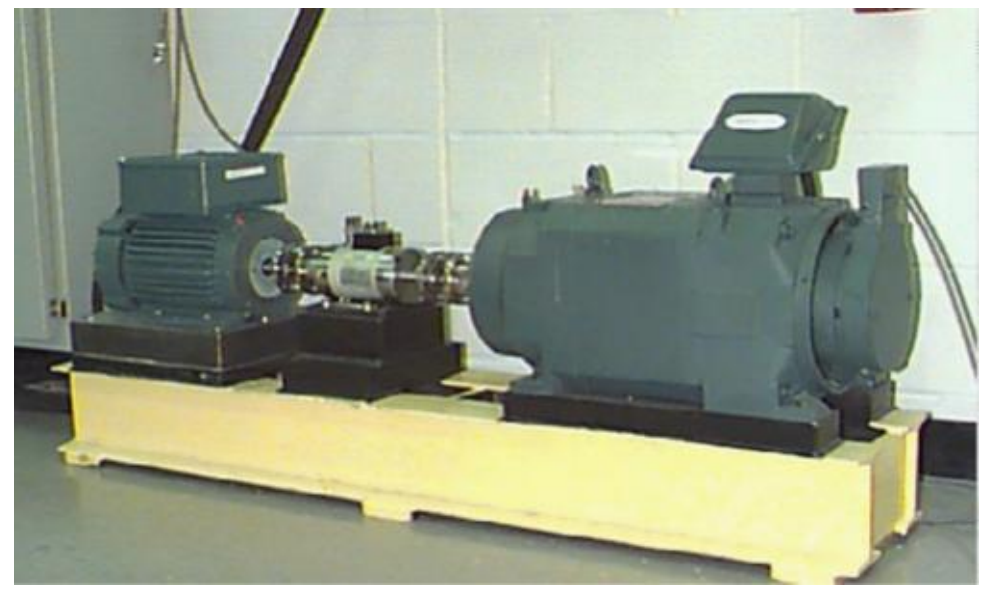

*Figure 11: Experimental Setup of the CWRU Dataset [36].*

### 3.2. Gas Turbine Industrial Dataset

In this study, vibration data from an industrial MGT-40 gas turbine with a rated capacity of 42.2 MW, used in a power-generation unit, were used [37]. Figure 12 shows an overview of the MGT-40 gas turbine under study. The data were recorded by the turbine and generator vibration monitoring system and include measurements from 12 vibration sensors installed at various locations on the turbine-generator assembly, sampled at 25 kHz. Based on operating reports and maintenance records, the data were divided into three distinct operational conditions. The first category corresponds to the period when the equipment was affected by the imbalance phenomenon, and the vibration symptoms of this condition were observed in the signals. This portion of the data was designated "Before Imbalance" and represents the conditions prior to the implementation of corrective actions.

After the maintenance team identified the source of the imbalance, the troubleshooting and corrective adjustments process began. The data recorded during this transitional period, where the system's dynamic behavior was changing and had not yet reached a stable state, were classified as the During Imbalance class. Finally, after the corrective actions were completed and the equipment's return to optimal operating conditions was confirmed, new data were recorded, representing the stable state post-imbalance correction. They were used as the After Imbalance class. This segmentation enables the investigation of the gradual changes in the turbine's vibrational behavior from the faulty state to normal conditions, transforming the detection problem into a three-class classification based on the different stages of the fault's evolution.

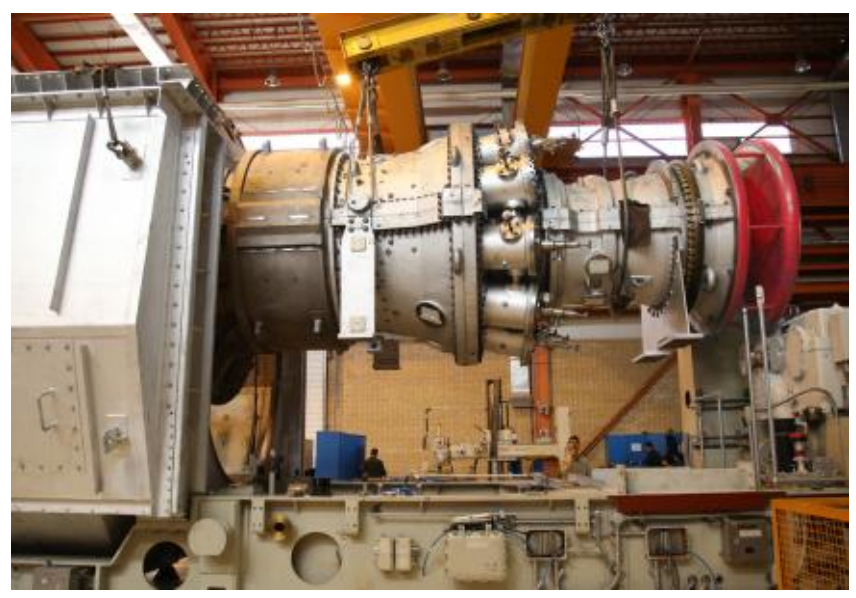

*Figure 12: Industrial MGT-40 Gas Turbine Used in This Study [38].*

## 4. Results

In this study, the CWRU dataset was used as the source domain and the gas turbine dataset as the target domain. The source dataset comprises 2545 samples across 10 classes, and the target dataset comprises 2982 samples across 3 classes. To evaluate the model's performance, the target domain data were split into training and testing sets in an 80:20 ratio. Additionally, during the training phase, data augmentation techniques – including random resized crop, color jitter, and random horizontal flip – were employed to increase data diversity and improve the model's generalization capability.

The main backbone of the proposed model was designed based on ConvNeXtV2-Femto. Compared with the larger versions of the ConvNeXtV2 family, this version has lower computational complexity and fewer parameters. The feature maps extracted at the different stages of the encoder have 48, 96, 192, and 384 channels, respectively, and in the final configuration, 7,191,759 of the model's 7,415,871 parameters were trainable. Initial results showed that ConvNeXtV2-Femto provides adequate performance on the studied defect diagnosis task while maintaining a low computational cost; therefore, it was selected as the final backbone.

### 4.1. Metrics Table & Confusion Matrix

In order to examine the effectiveness of the developed model, it was compared with various well-known deep learning frameworks. They include ResNet50, DenseNet121, EfficientNet-B0, Xception, Swin Transformer, and ConvNeXtV1. All the mentioned models were trained and tested under identical conditions to ensure a fair comparison. The following metrics have been utilized to test the performance: Accuracy, Precision, Recall, and F1-Score.

Table 1 presents the results of this comparison. As shown, the proposed model has demonstrated better performance than the others in the target data. This indicates that the combination of the multi-domain signal representation, the MDCA mechanism, and the multi-task learning strategy has extracted more discriminative features. As a result, the proposed method improved diagnostic accuracy in cross-domain knowledge transfer scenarios.

The test-set confusion matrix is shown in Figure 13 to examine the model's performance across classes. The largest number of samples is on the main diagonal of the matrix, leading to high classification accuracy. The number of errors recorded outside the main diagonal is very limited. In fact, only a small number of samples have been misclassified across classes. Moreover, none of the classes has experienced a significant performance drop, and no clear pattern of systematic errors is observed. This reveals that the features extracted by the model were able to establish suitable boundaries between the different classes.

*Table 1: Performance Comparison of Different Deep Learning Models.*

| Model | Accuracy | Precision | Recall | F1-Score |
|---|---|---|---|---|
| ResNet50 | 95.97 | 96.22 | 95.58 | 95.87 |
| DenseNet121 | 96.81 | 96.87 | 96.81 | 96.84 |
| EfficientNet-B0 | 90.60 | 90.98 | 90.14 | 90.50 |
| Xception | 93.96 | 94.87 | 93.12 | 93.85 |
| Swin Transformer | 94.46 | 94.81 | 93.99 | 94.37 |
| ConvNeXtV1 | 97.32 | 97.38 | 97.32 | 97.31 |
| Proposed Method | **98.83** | **99.02** | **98.72** | **98.81** |

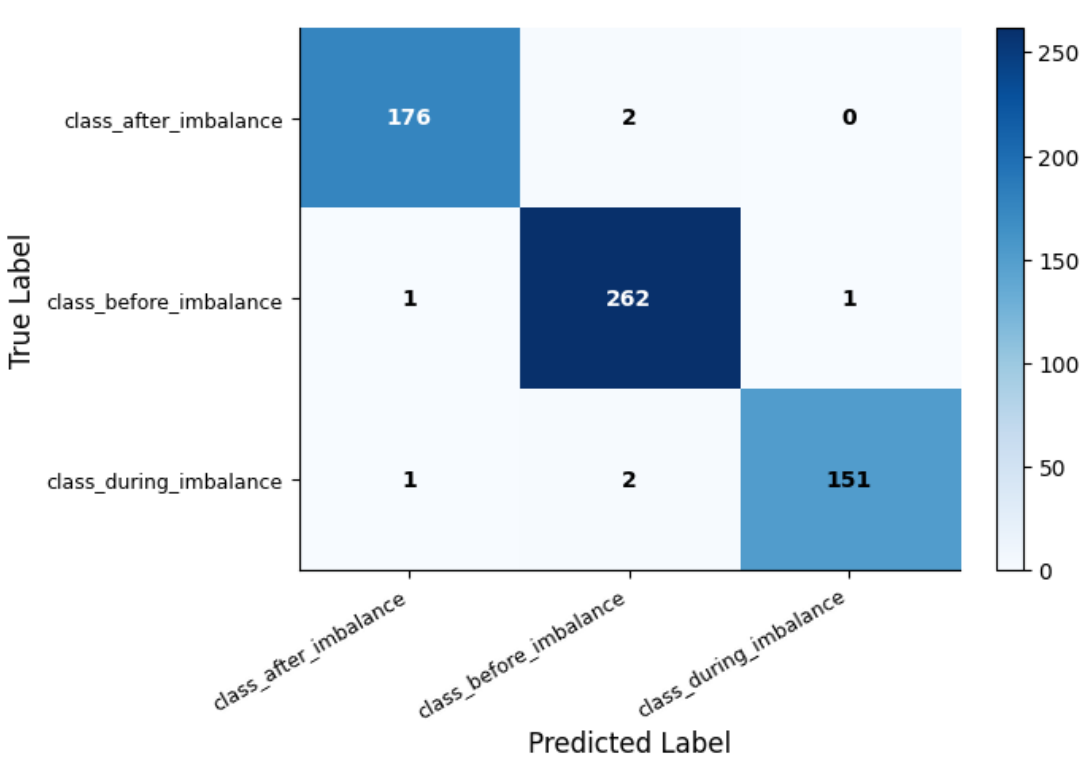

*Figure 13: Confusion Matrix on the Test Set.*

### 4.2. Training Curves

The training and validation curves for the model are depicted in Figure 14. During training on the source domain, the loss initially decreased considerably and then stabilized after some epochs. Furthermore, the training and validation accuracy showed a rise with time. From the proximity between the training and validation curves in this case, it can be understood that the model has been trained successfully.

A similar trend is observed during the fine-tuning phase on the target domain. The gradual decrease in loss and the steady increase in accuracy indicate stable knowledge transfer from the source to the target domain. Moreover, the small gap between the training and test curves across most epochs indicates that the model generalizes well to unseen data.

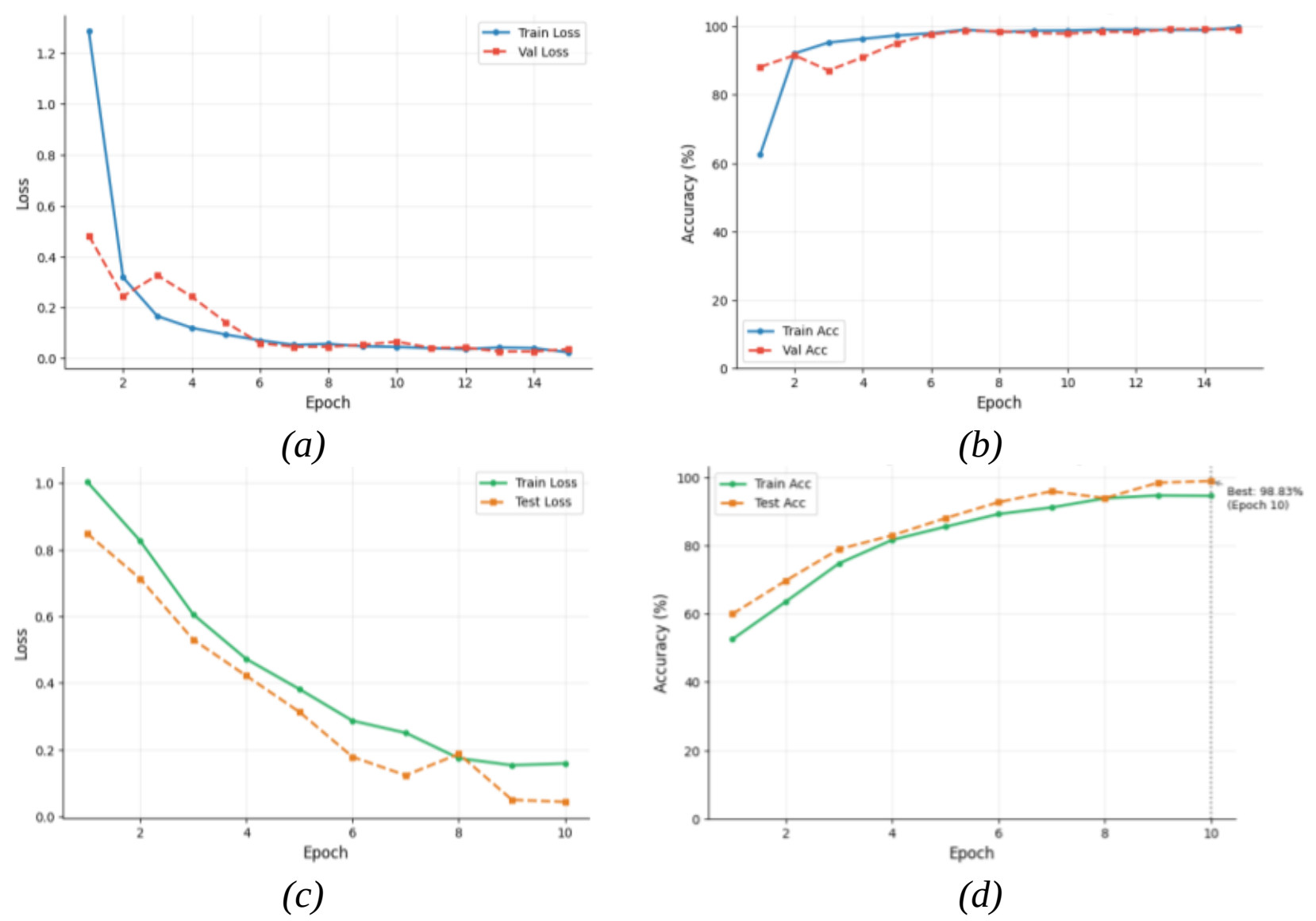

*Figure 14: Training History of the Proposed XMDCA-TL Model: (a) Source-Domain Loss, (b) Source-Domain Accuracy, (c) Target-Domain Loss, and (d) Target-Domain Accuracy.*

### 4.3. Ablation Study

To evaluate the contribution of each component of the proposed architecture, several simplified versions of the model were created, and their performance was compared with that of the full model. The results of these experiments are presented in Figure 15. As shown, the full model achieved the highest performance, whereas removing certain components resulted in a significant drop in accuracy.

The greatest performance drop was observed when the MDCA module and the transfer learning mechanism were removed. This result indicates that the model heavily relies on learning interactions among different signal domains and on transferring knowledge from the source domain to the target domain. In contrast, removing the denoising decoder caused only a limited decrease in accuracy. Although this component's impact on the final performance is less than that of other components, the results show that simultaneously reconstructing and learning the data structure still improves the quality of the extracted representations. Overall, the results of this study indicate that the model's final performance is the result of the collaboration of all architectural components, but the MDCA and transfer learning played the most significant role in achieving the final accuracy.

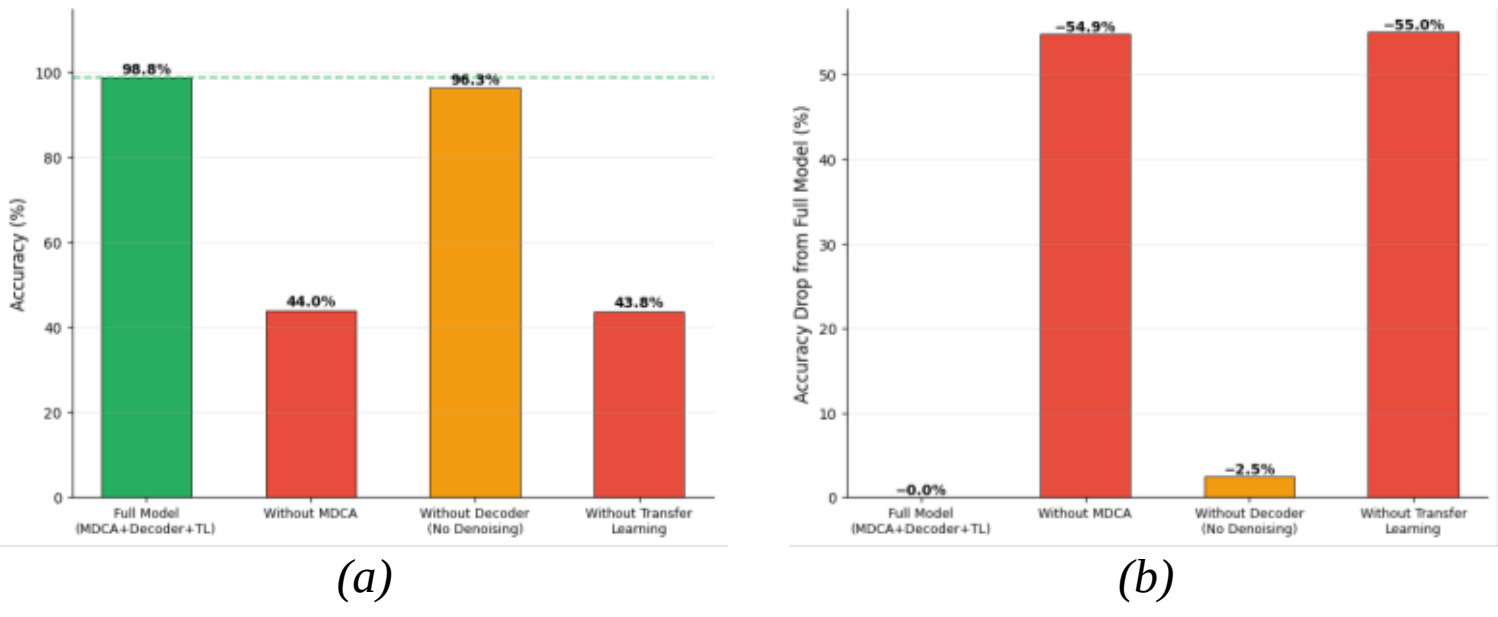

*Figure 15: Ablation Study Results: (a) Classification Accuracy of Model Variants and (b) Contribution of Individual Components.*

### 4.4. UMAP & t-SNE Analysis

To examine the quality of the extracted features and the distribution of data in feature space, two dimensionality reduction methods, t-SNE and UMAP, were used. The results of the two-dimensional visualization of the encoder's output features after applying the MDCA module are presented in Figure 16. As can be seen, the samples corresponding to the different source classes have, in most cases, formed separate, compact clusters, indicating the model's ability to extract discriminative features.

On the other hand, the samples from the target domain are also located in specific regions of the feature space and are closer to some of the source clusters. This indicates that knowledge transferred from the CWRU data has created a suitable structure for representing the target data. Moreover, the results of both t-SNE and UMAP show a similar trend, where the boundaries between many classes are preserved, and limited overlap is observed among the clusters. Overall, this analysis indicates that the features generated by the encoder and the MDCA module have adequate discriminative power and have reduced the distance between the source and target domains.

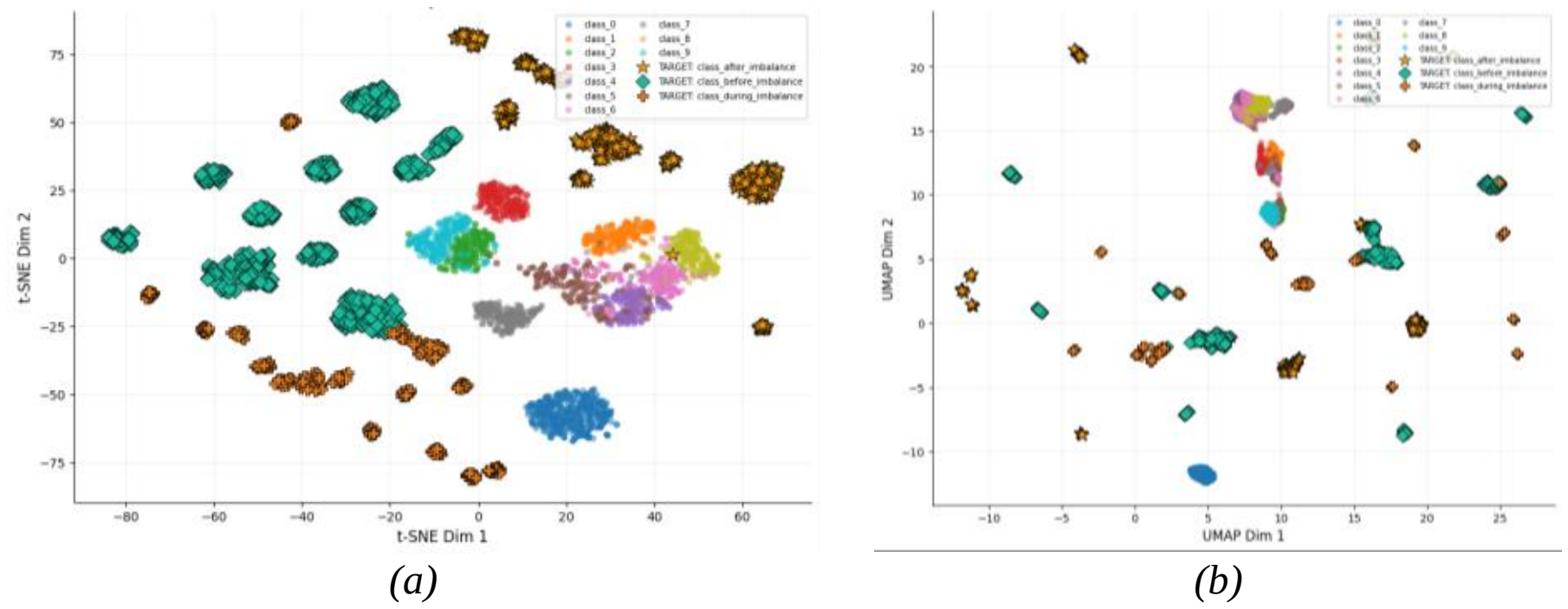

*(a)* *(b)*

*Figure 16: Feature Space Visualization of the Extracted Representations after MDCA: (a) t-SNE Projection and (b) UMAP Projection.*

### 4.5. Noise Robustness

To evaluate the model's stability under different environmental conditions, its performance was examined across varying noise levels. For this purpose, noise with varying SNRs was added to the target data, and the Precision, Recall, and F1-Score metrics were then calculated. The results of this experiment are shown in Figure 17.

As shown, the highest performance is observed for noise-free data. As the SNR ratio decreases, the model's performance degrades, with the greatest reduction observed at SNR=0 dB. However, as the SNR increases from 0 to 5, 10, and 15 decibels, improvements in the metrics are clearly visible. This behavior indicates that the proposed model effectively preserves defect-related information under noisy conditions and that the extracted features are reasonably stable against environmental disturbances. In addition, the results confirm the role of the reconstructive decoder and the denoising process in enhancing the model's resilience to data corruption and degradation of signal quality.

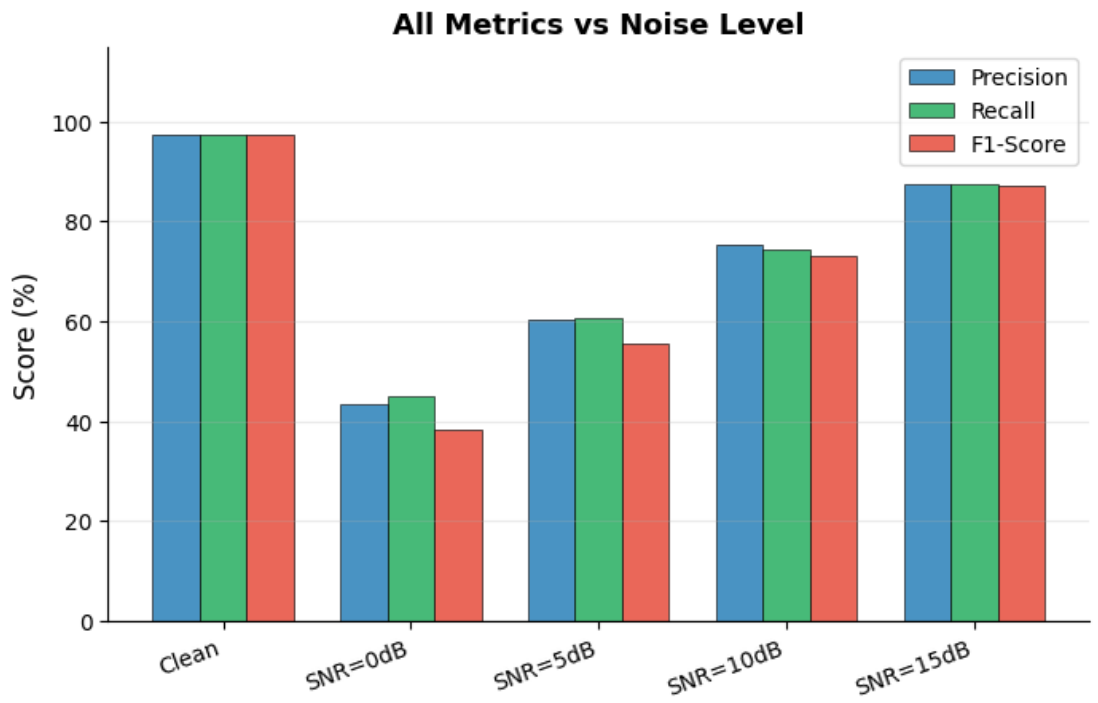

*Figure 17: Performance Evaluation Under Different Noise Levels.*

### 4.6. Denoising Reconstruction Analysis

To evaluate the performance of the denoising branch, several reconstructed samples are shown in Figure 18. As shown, the model has recovered a significant portion of the removed information while preserving the image's overall structure in both donut masking and patch masking modes.

A qualitative inspection of the results shows that the main patterns across the RGB channels remain discernible after reconstruction. This indicates that the encoder not only learns the features required for classification but also preserves structural information in the data's latent representations. Additionally, the ability to reconstruct the removed regions shows that the model has learned the interactions among the time, frequency, and time-frequency domains and uses this information to estimate the missing parts. This feature can play a crucial role in enhancing the model's robustness against noise, incomplete data, and domain shifts.

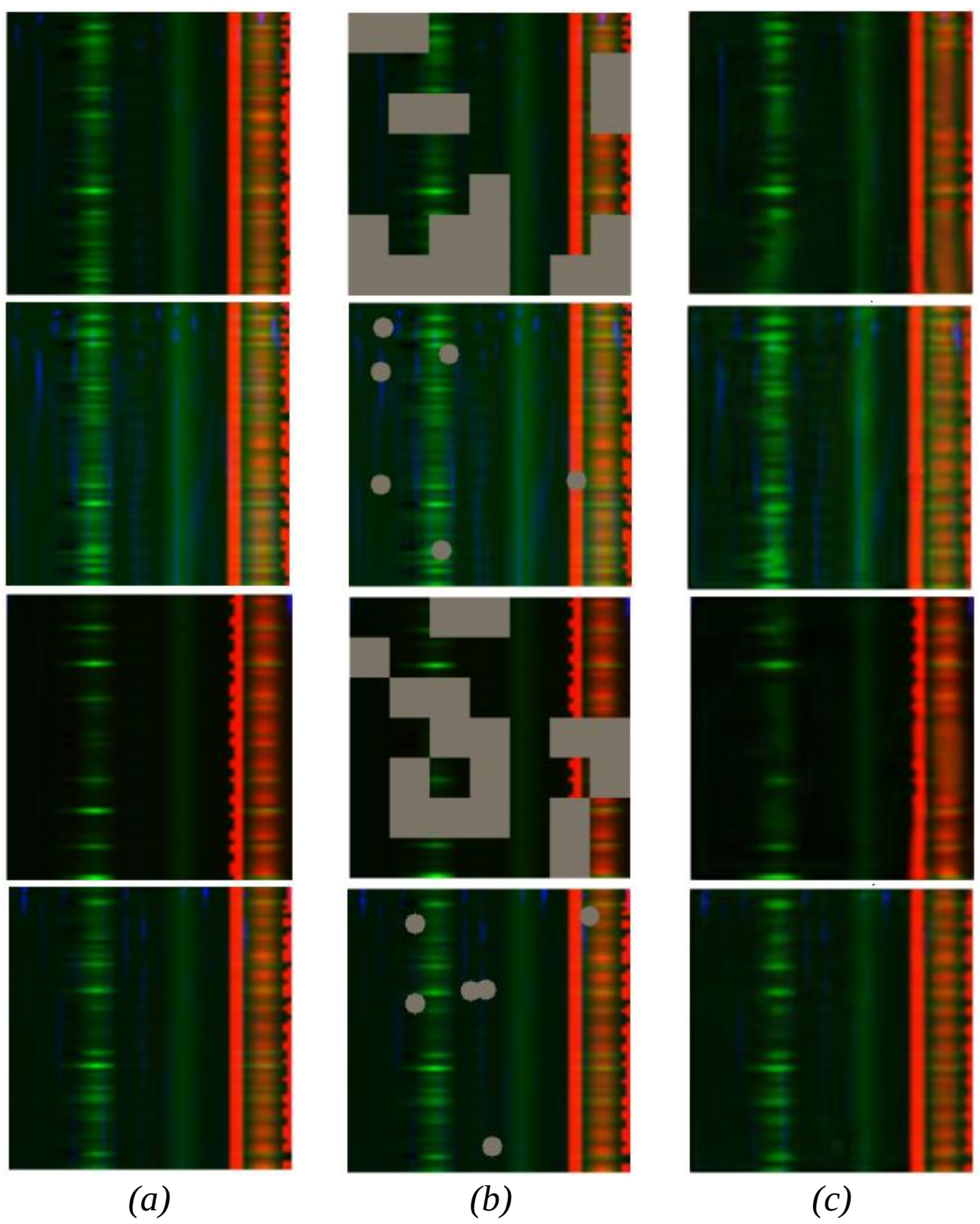

*Figure 18: Denoising Reconstruction Examples: (a) Original, (b) Masked, and (c) Reconstructed Images.*

### 4.7. XAI Analysis

To examine the model's decision-making process, the Grad-CAM++ method was used to generate heat maps. Figure 19 shows several examples of the input images, importance maps, and the model's final output for different classes. Areas with warmer colors indicate the parts that contributed more to the network's decision-making, while the blue-colored areas had less influence on the model's output. The model for all classes has focused on specific parts of the image and has not based its decision on the entire image. Furthermore, the extracted attention patterns are not identical across samples, and the important regions vary by class. This indicates that the network has identified the distinguishing features of each class and used them for classification.

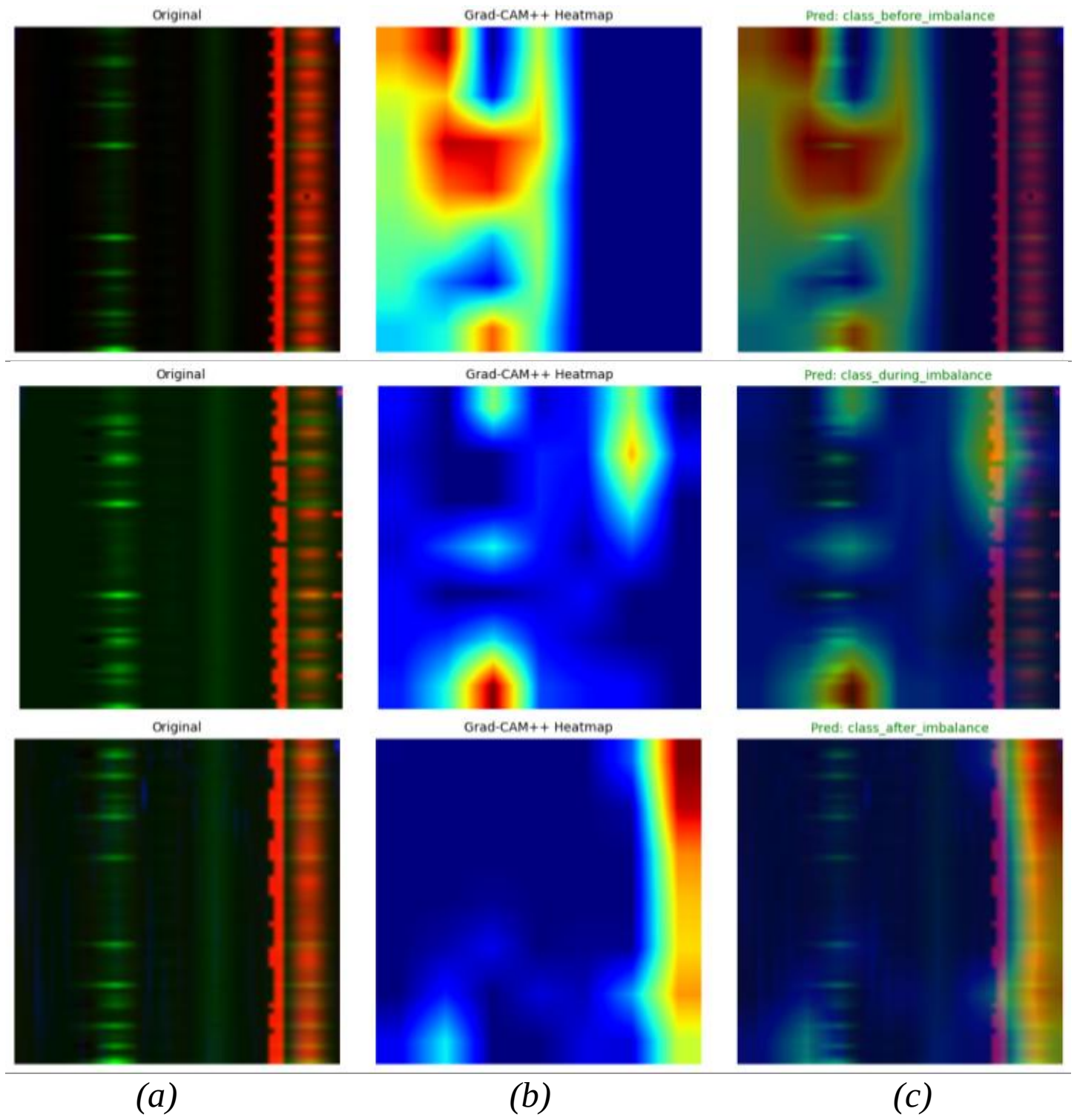

*Figure 19: Grad-CAM++ Analysis Results: (a) Original Images, (b) Importance Maps, and (c) Model Prediction.*

The average Grad-CAM++ maps for each class were calculated and shown in Figure 20 to gain an overall view of the model's behavior. The results depict that the model's attention pattern differs across the three classes, and each class has its own characteristic regions. Notably, the before imbalance class exhibits important regions that are not confined to a single, limited part of the image. This behavior is consistent with the nature of this class. In the initial stages of imbalance, changes typically appear gradually and diffusely across different parts of the signal, rather than being confined to a specific local. In contrast, the other two classes show a greater focus on specific areas of the image, indicating the presence of more distinct, local features for classification.

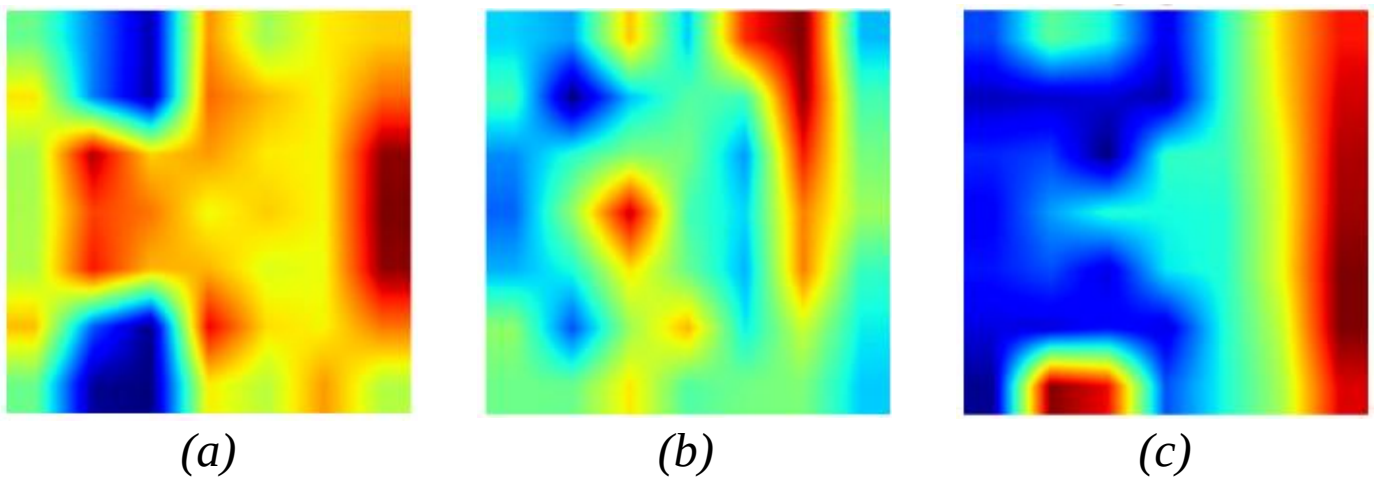

(a) (b) (c)

*Figure 20: Average Grad-CAM++ Maps for Different Classes: (a) Before Imbalance, (b) During Imbalance, and (c) After Imbalance.*

The results of the Grad-CAM++ reverse mapping show that the model focuses on different parts of the signal for each class. As shown in Figure 21, in the before imbalance class, high-importance regions are distributed across several distinct time intervals. In other words, the model's focus is not limited to a single part of the signal. In contrast, in other classes, high-importance regions appear more frequently within specific, concentrated time intervals. In the during imbalance example, the highest importance is observed in the middle section of the signal. As a matter of fact, features related to the imbalance event are more prominent during this time frame. Also, in the after imbalance class, the model's focus is primarily on the beginning of the first half of the signal, after which the importance value gradually decreases. This indicates that certain vibration patterns appear within specific time intervals of the signal and play a decisive role in the classification process. Overall, the results show that the model can identify the parts of the signal that contain the most diagnostic information from a physical perspective, thereby confirming the interpretability of the model's decisions.

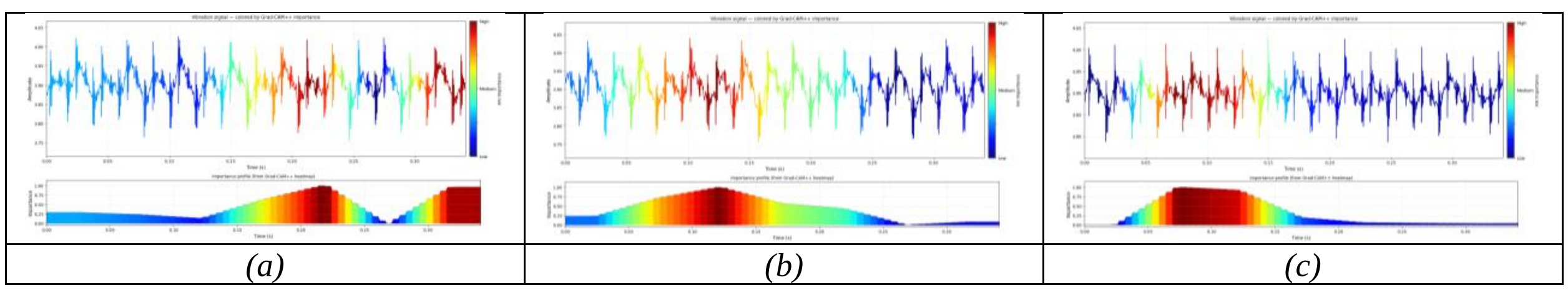

| (a) | (b) | (c) |
|---|---|---|

*Figure 21: Time-Domain Interpretation Results Using Grad-CAM++ Reverse Mapping: (a) Before Imbalance, (b) During Imbalance, and (c) After Imbalance.*

To better understand how the model leverages multi-domain information, the behavior of the MDCA module was examined from three different perspectives. Figure 22 shows the overall importance of each domain in the two MDCA stages. It is observed that in stage 3 of the backbone, almost all feature energy is concentrated in the frequency domain, while in stage 4, the time domain's share increases significantly, becoming the most effective source of information. Since the output of each branch is the result of the cross-attention process, these results do not simply indicate the independent use of a single domain. In other words, when the time domain achieves the highest importance in stage 4, part of that importance stems from the complementary information it has received from the frequency domain. The disaggregated results for each class in Figure 23 also confirm this trend. In stage 3, for all three classes, the frequency range accounts for the largest share. However, in stage 4, the attention distribution pattern becomes more balanced, and the time range takes on the greatest importance.

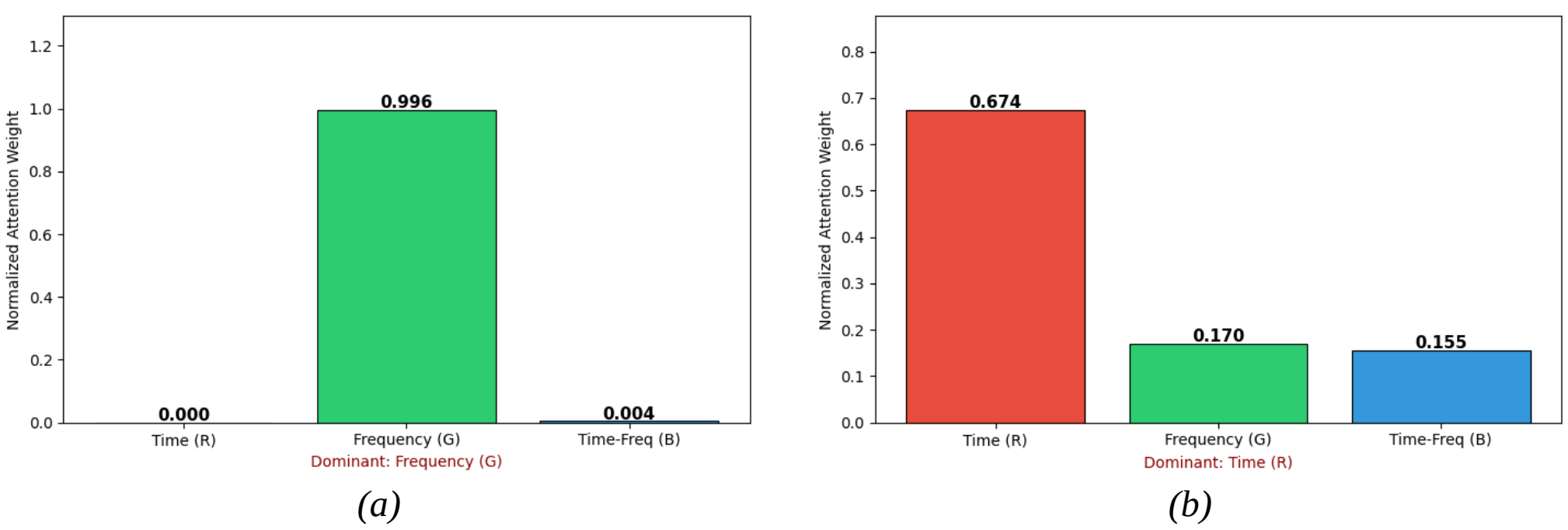

*(a)* *(b)*

*Figure 22: Domain Importance Analysis in MDCA Stages: (a) Stage 3 and (b) Stage 4 of the Backbone.*

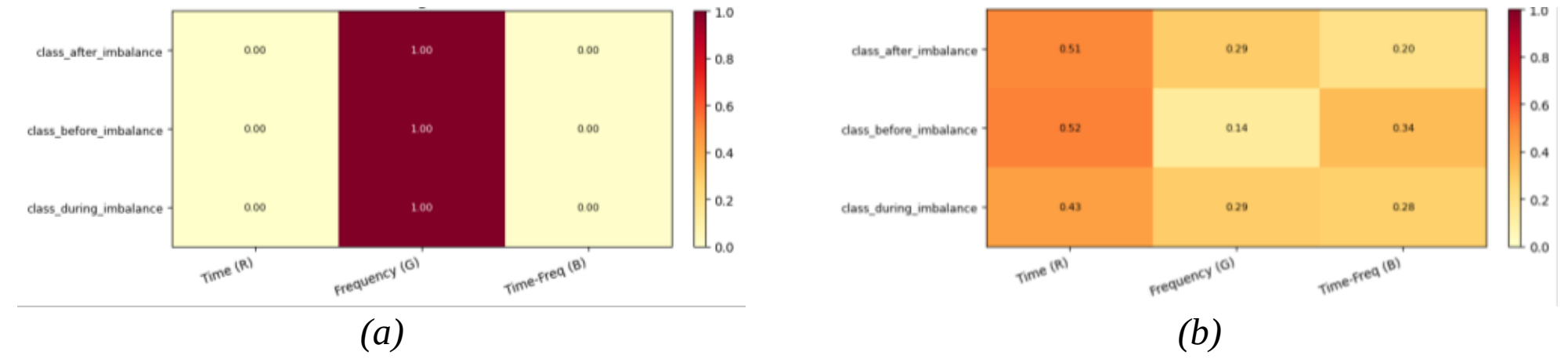

*(a)* *(b)*

*Figure 23: Domain Importance Across Different Classes: (a) Stage 3 and (b) Stage 4 of the Backbone.*

In the following overall analysis of the MDCA module, Figure 24 shows the dependence of each fault class on the three information domains separately. It is shown that across all three classes, the time domain has the largest share, but the pattern of attention distribution among the domains differs. In the after imbalance class, the time domain's share reaches about 0.69, indicating that the diagnosis of this condition is primarily based on temporal changes in the signal. In contrast, in the before imbalance class, the time-frequency domain's share increases to about 0.31, suggesting the presence of transient patterns and concurrent changes in both time and frequency prior to the occurrence of imbalance. Also, in the during imbalance class, although the time domain remains dominant, the increased share of the frequency domain compared to other classes indicates that frequency components also play an effective role in identifying this condition.

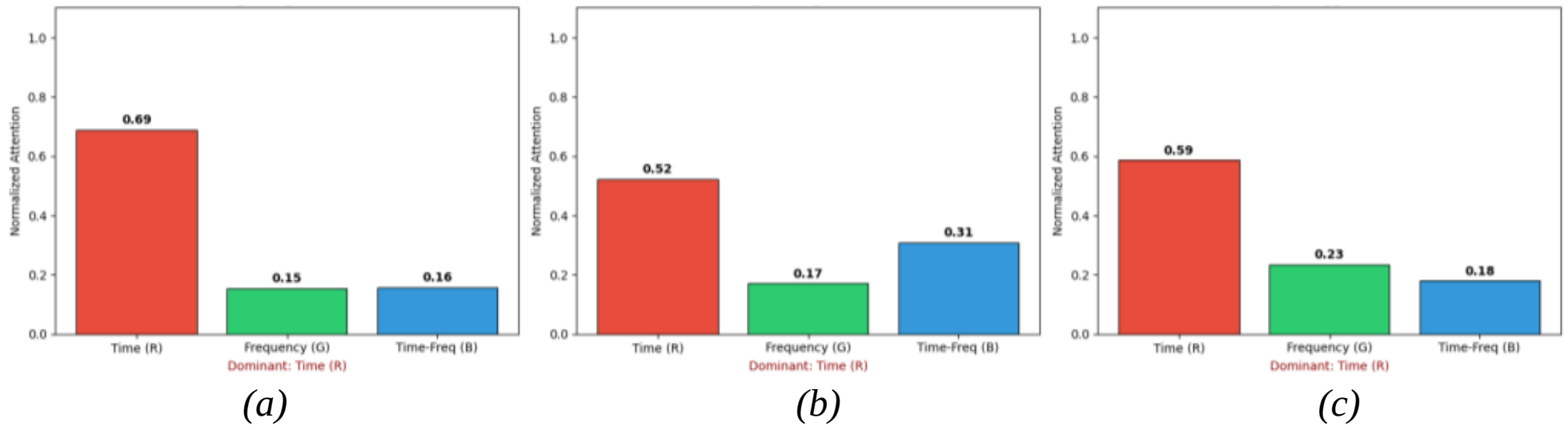

*(a)* *(b)* *(c)*

*Figure 24: Fault Signatures Across Different Information Domains: (a) After Imbalance, (b) Before Imbalance, and (c) During Imbalance.*

To better understand how information is exchanged among the three domains, the cross-domain attention flow generated in the MDCA module was examined in Figure 25. In all three classes, the Time ← Frequency path receives the greatest attention. This result indicates that frequency information is crucial for enriching temporal representations. In contrast, the Frequency ← Time-Frequency path had the smallest contribution in most classes. In fact, time-frequency information serves more as a supplement than as the primary source of feature extraction. The Time-Frequency ← Time path also shows a relatively high value, especially in the before imbalance class. These observations are also consistent with the attention maps in Figure 26. Each path focuses on different regions of the image, indicating that MDCA, rather than using each domain independently, establishes a directed information exchange process among all domains.

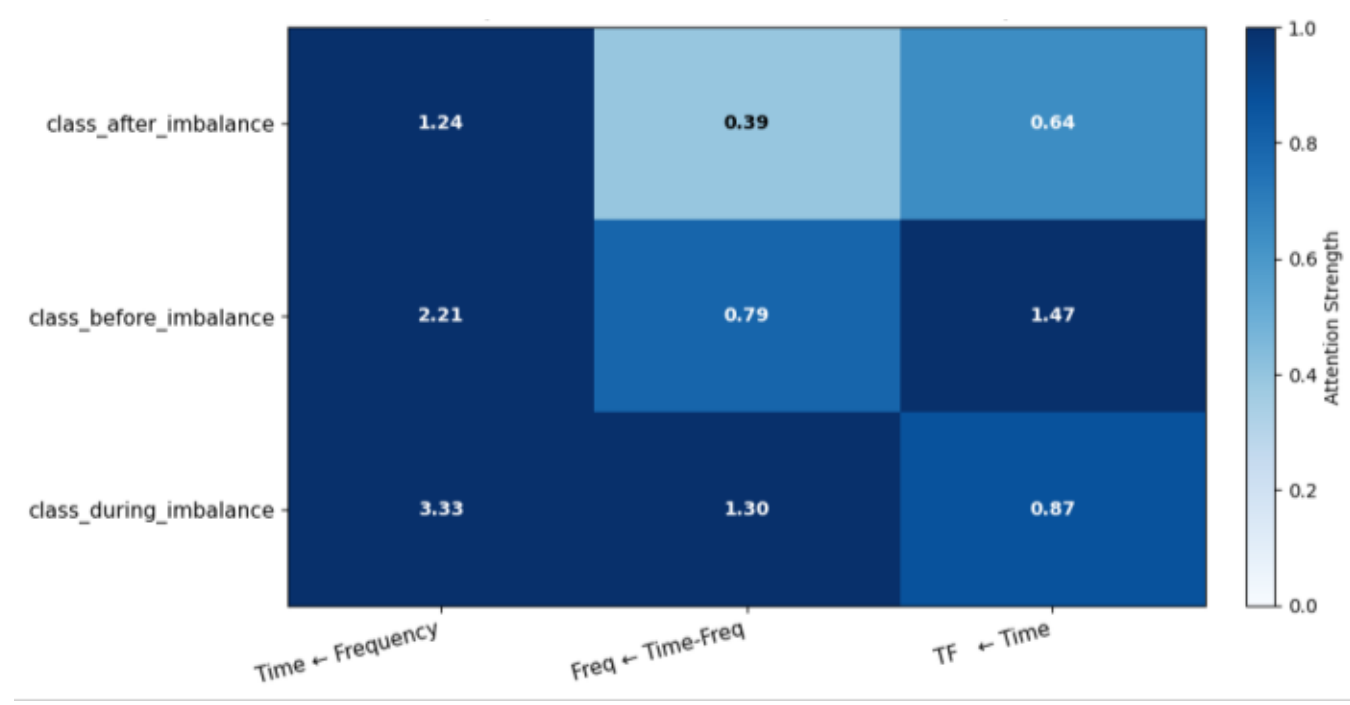

*Figure 25: Cross-Domain Attention Flow Across Different Fault Classes.*

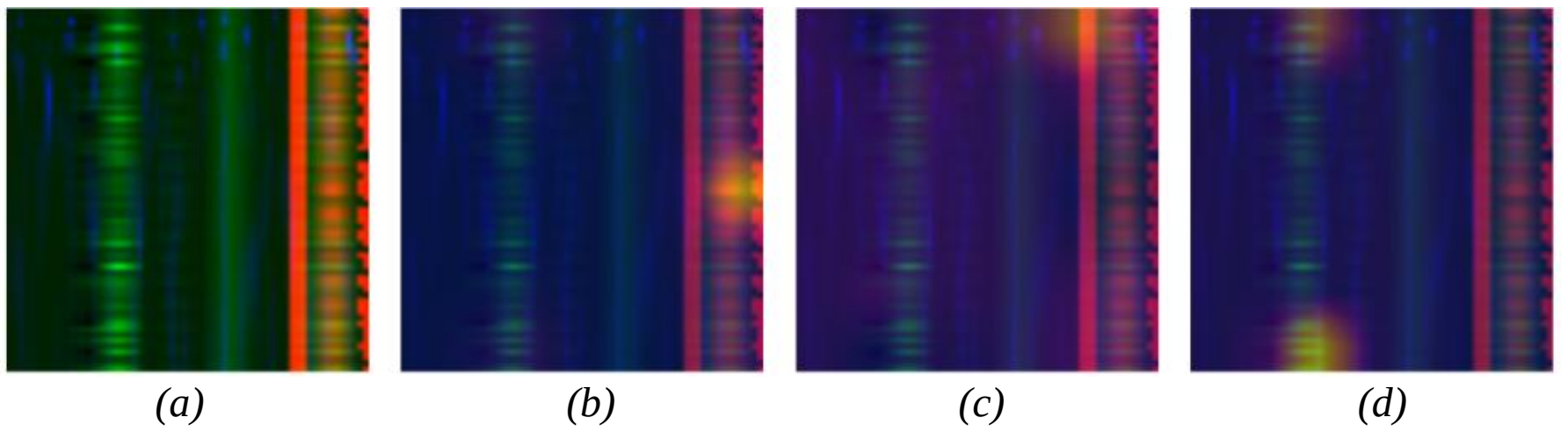

*(a) (b) (c) (d)*

*Figure 26: Cross-Domain Attention Map: (a) Original Image, (b) Time ⟵ Frequency, (c) Frequency ⟵ Time-Frequency, and (d) Time-Frequency ⟵ Time.*

To examine the effect of knowledge transfer, the distributions of domain importance in the initial and adapted models were first compared. As shown in Figure 27, after the adaptation process, the share of the time domain increased from 0.30 to 0.55. In contrast, the importance of the frequency and time-frequency domains decreased from 0.33 to 0.20 and from 0.37 to 0.24, respectively. This change indicates that after adapting to the target domain, the model has become more dependent on temporal patterns and has represented part of the frequency information more effectively in temporal representations. The results in Figure 28 also confirm this behavior. In the initial model, the Grad-CAM++ map focuses on broader, more scattered regions of the image. Still, after adaptation, the model's attention is confined to more specific areas and key structures related to the defect. The difference image also shows that the greatest changes in attention occurred in the same regions that became more decisive in the model's decision-making after knowledge transfer. Overall, these results indicate that the fine-tuning process has effectively represented part of the frequency and time-frequency information in more effective temporal representations.

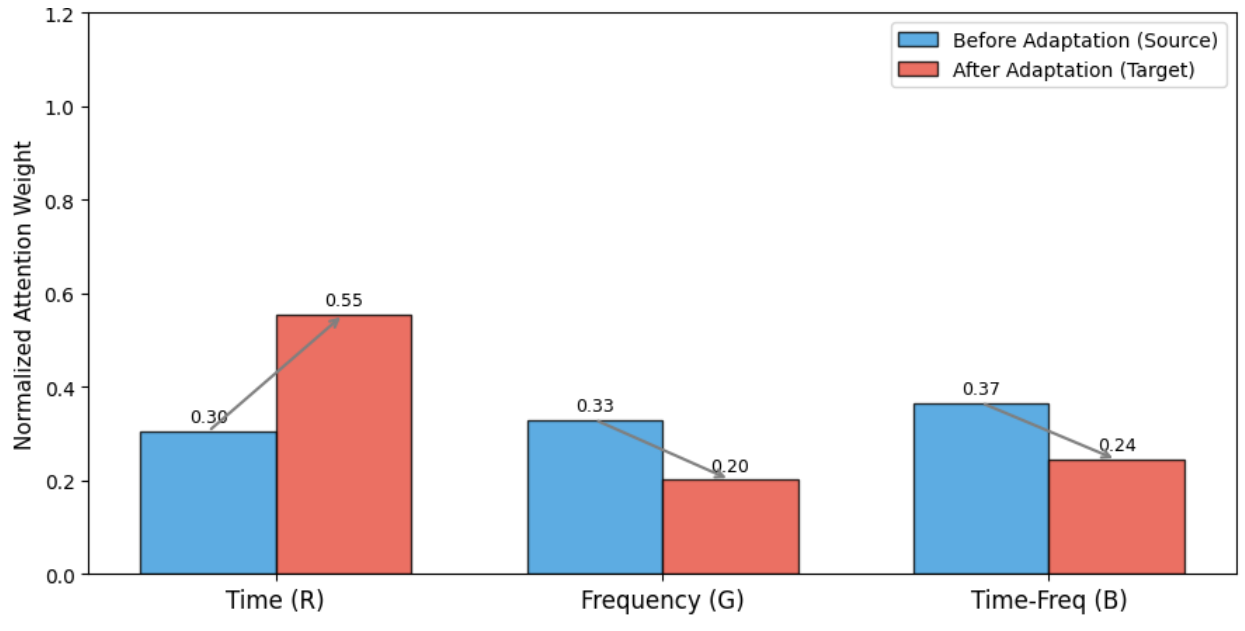

*Figure 27: Transfer Gap and Domain Importance: Before vs After Adaption.*

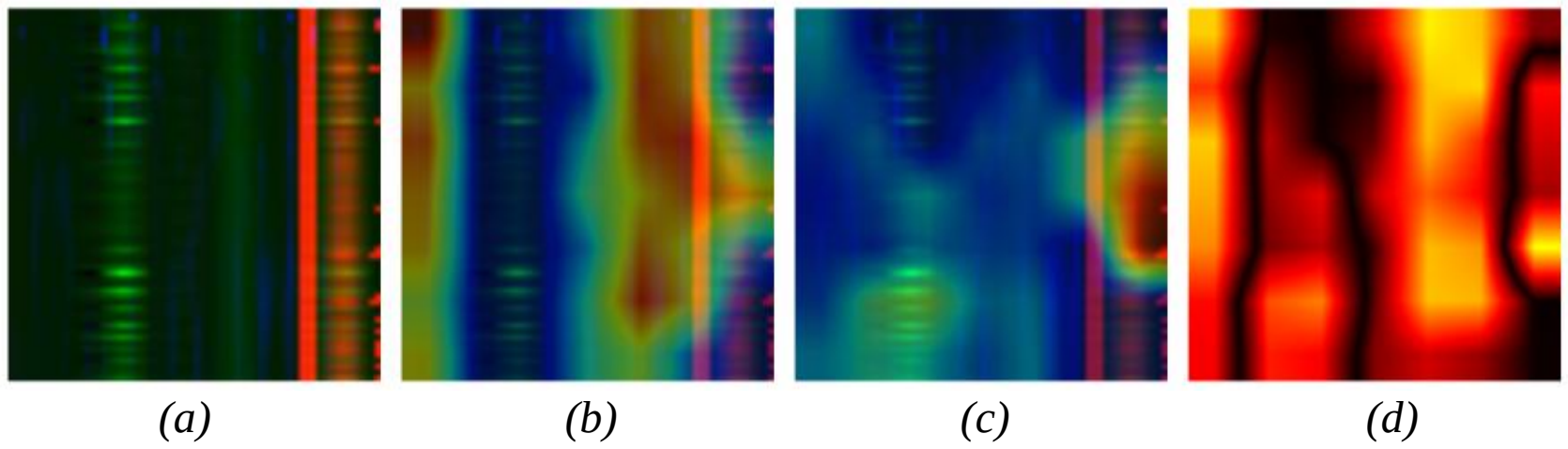

*(a)* *(b)* *(c)* *(d)*

*Figure 28: Grad-CAM++ Comparison:(a) Original Image, (b) CAM (Source Model), (c) CAM (Adapted Model), and (d) Difference.*

To examine the degree of uncertainty during reconstruction, the difference between the input and reconstructed output images was calculated in all three domains. As shown in Figure 29, the largest reconstruction error across all classes occurs in the time domain, while the frequency domain has the lowest error. This indicates that a significant portion of the data's unstable and noisy components was concentrated in the temporal domain. Hence, the decoder applied more changes there. In addition, Figure 30 shows that the regions of uncertainty are mainly concentrated in the temporal sections, while the main patterns related to the corruption are preserved after reconstruction.

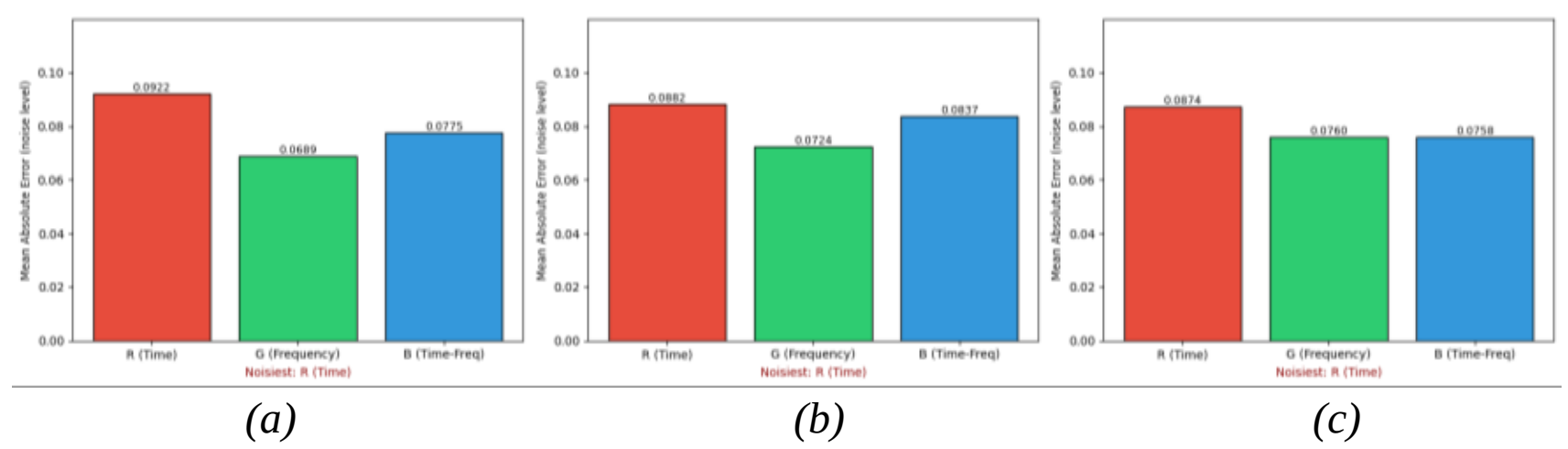

*(a)* *(b)* *(c)*

*Figure 29: Denoising Uncertainty Across Different Classes: (a) After Imbalance, (b) Before Imbalance, and (c) During Imbalance.*

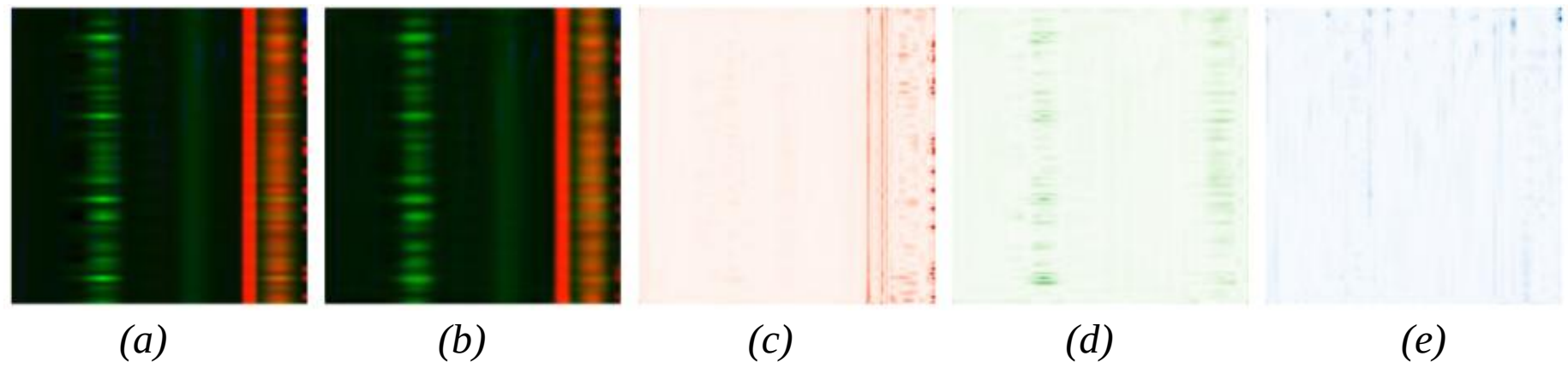

*(a)* *(b)* *(c)* *(d)* *(e)*

*Figure 30: Denoising Uncertainty Maps: (a) Original, (b) Denoised, (c) Time Domain Noise, (d) Frequency Domain Noise, and (e) Time-Frequency Domain Noise.*

## 5. Conclusion

In this study, the XMDCA-TL framework was presented as a multi-task, explainable transfer learning method for fault diagnosis in industrial gas turbines. In this framework, vibration signals were fed into the model using a multi-domain RGB representation spanning the time, frequency, and time-frequency domains, and a ConvNeXtV2-based encoder with an MDCA mechanism was employed to extract features and model interactions across domains. Also, to improve the quality of the learned representations and increase the model's robustness to noise and domain shifts, a self-supervised strategy based on hybrid masking and a UNet-based decoder was designed to reconstruct the masked regions. The results from knowledge transfer from laboratory data to real-world MGT-40 gas turbine data showed that the proposed method can identify various imbalance patterns in real-world industrial conditions with 98.83% accuracy. Furthermore, UMAP and t-SNE analyses, as well as noise robustness tests, demonstrated the model's ability to learn separable and stable representations. Results from XAI analyses also showed that the model's decisions were based on significant signal regions, with temporal information making the greatest contribution to fault diagnosis. In contrast, frequency and time-frequency information played a complementary role in improving representation quality. These results indicate that XMDCA-TL can be used as an effective and reliable solution for intelligent condition monitoring and predictive maintenance systems in industrial rotating machinery.

## CRediT authorship contribution statement

**Amir Jahangard Takaloo:** Conceptualization, Investigation, Methodology, Software, Validation, Visualization, Writing – original draft, Writing – review & editing. **Mahdi Aliyari Shoorehdeli:** Conceptualization, Investigation, Methodology, Supervision, Visualization, Writing – review & editing. **Ehsan Mohammadi:** Conceptualization, Investigation, Supervision, Writing – review & editing.

## Declaration of Interest Statement

The authors declare that they have no known competing financial interests or personal relationships that could have influenced the work reported in this paper.

## Acknowledgments

This research did not receive any specific grant from funding agencies in the public, commercial, or not-for-profit sectors.